\documentclass[a4paper,12pt]{article}
\usepackage{amsmath}
\usepackage{amssymb}
\usepackage[dvipdfmx]{graphicx}
\usepackage{bm}

\usepackage{wrapfig}
\usepackage[nosort]{cite}

\newcommand{\nn}{\nonumber}
\newcommand{\tr}{\text{tr}\,}

\newcommand{\vev}[1]{\left\langle #1 \right\rangle}

\newcommand{\cO}{{\cal O}}
\newcommand{\binomi}[2]{\begin{pmatrix} #1 \\ #2 \end{pmatrix}}
\newcommand{\Ai}{\text{Ai}}
 \def\XXint#1#2#3{{\setbox0=\hbox{$#1{#2#3}{\int}$}
 \vcenter{\hbox{$#2#3$}}\kern-.5\wd0}}

\begin{document}
\thispagestyle{empty} \addtocounter{page}{-1}
\vspace*{1cm}

\begin{center}
{\large \bf Two-point functions at arbitrary genus 
and its resurgence structure 
in a matrix model for 2D type IIA superstrings}\\
\vspace*{2cm}
Tsunehide Kuroki$^{*,\ddagger}$\\
\vskip0.7cm
{}$^*${\it Theoretical Physics Laboratory, Toyota Technological Institute}\\
\vspace*{1mm}
{\it 2-12-1 Hisakata, Tempaku-ku, Nagoya 468-8511, Japan}\\
\vspace*{0.2cm}
{\tt kuroki@toyota-ti.ac.jp}\\
\end{center}
\vskip1.5cm
\centerline{\bf Abstract}
\vspace*{0.3cm}
{\small 
In the previous papers, it is pointed out that a supersymmetric double-well matrix model 
corresponds to a two-dimensional type IIA superstring theory 
on a Ramond-Ramond background at the level of correlation functions. 
This was confirmed by agreement between their planar correlation functions. 
The supersymmetry in the matrix model corresponds to the target space 
supersymmetry and it is shown to be spontaneously broken by nonperturbative effect. 
Furthermore, in the matrix model we computed one-point functions 
of single-trace operators to all order of genus expansion in its double scaling limit. 
We found that this expansion is stringy and not Borel summable and hence 
there arises an ambiguity in applying the Borel resummation technique. 
We confirmed that resurgence works here, 
namely this ambiguity in perturbative series in a zero-instanton sector 
is exactly canceled by another ambiguity in a one-instanton sector 
obtained by instanton calculation. 
In this paper we extend this analysis and study resurgence structure 
of the two-point functions of the single trace operators. 
By using results in the random matrix theory, we derive two-point functions 
at arbitrary genus and see that the perturbative series 
in the zero-instanton sector again has an ambiguity. 
We find that the two-point functions inevitably have logarithmic singularity 
even at higher genus. In this derivation we obtain a new result 
of the two-point function expressed by the one-point function 
at the leading order in the soft-edge scaling limit of the random matrix theory. 
We also compute an ambiguity in the one-instanton sector by using the Airy kernel, 
and confirm that ambiguities in both sectors cancel each other 
at the leading order in the double scaling limit. We thus clarify resurgence structure 
of the two-point functions in the supersymmetric double-well matrix model.  

}
\vspace*{1.1cm}

\newpage

%\maketitle  IS IGNORED %%%%%%%%%%%
%%%%%%%%%%%%%%%%%%%%%%%%%%%%%%%%%%%%%%%%%%%%%%%%%%%%%%%%%%%%%%%%%%%%%%%%%%%%%%
\section{Introduction}
\label{sec:intro}
\setcounter{equation}{0}
%%%%%%%%%%%%%%%%%%%%%%%%%%%%%%%%%%%%%%%%%%%%%%%%%%%%%%%%%%%%%%%%%%%%%%%%%%%%%%
So far string theory has been defined only perturbatively except peculiar cases 
with much higher symmetry or much fewer degrees of freedom. 
It is, however, understood that its nonperturbative effect is more relevant 
than in field theory and that they would play an important role in determining 
vacuum structure of string theory. 
 
On the other hand, resurgence 
\cite{Ec1,Pham1,BH1,Howls1,DH1,Costin1,Sauzin1,Sauzin2} has attracted 
lots of attention because it enables us to extract nonperturbative information
of our interest from data of higher-order perturbative expansion.\footnote
{Resurgence structure has been studied in various models and theories.  
See e.g. in quantum mechanics 
\cite{Alvarez1,ZinnJustin:2004ib,Dunne:2013ada,
Escobar-Ruiz:2015nsa,Misumi:2015dua,Behtash:2015zha,Gahramanov:2015yxk,
Fujimori:2016ljw,Dunne:2016jsr,Sulejmanpasic:2016fwr,Serone:2016qog,Basar:2017hpr,
Alvarez:2017sza}, 
string theories \cite{Marino:2008vx,Marino:2006hs,Grassi:2014cla} 
as well as quantum field theories \cite{Marino:2012zq,Aniceto:2015rua,
Dunne:2012ae,Cherman:2013yfa,Misumi:2014jua,Nitta:2014vpa,Behtash:2015kna,
Dunne:2015ywa,Buividovich:2015oju,Demulder:2016mja,Sulejmanpasic:2016llc,
Gukov:2016njj,Gang:2017hbs,Argyres:2012vv,Dunne:2015eoa,Yamazaki:2017ulc,
Russo:2012kj,Aniceto:2014hoa,Costin:2017ziv,Honda:2016mvg,Dorigoni:2017smz,Honda:2017cnz,
Fujimori:2018nvz,Ahmed:2018gbt,Dunne:2019aqp,Costin:2019xql}.
} 
Thus it is reasonable to expect that resurgence provides a powerful tool 
in analyzing nonperturbative aspects of string theory with respect to the 
string coupling constant. 
However, when we try to apply the resurgence to string theory, 
one of the biggest problems is difficulty in getting higher-genus contribution. 
In general, in order to obtain information on nonperturbative effect precisely, 
we need details of higher order in perturbative expansion, 
which are in general hard to deduce. 
In order to find perturbative expansion up to higher genus, we may consider theory 
with higher symmetry or without much degrees of freedom. Then we in turn have 
another problem that in such cases perturbative expansion itself often behaves well  
and that resurgence may not work to extract nonperturbative information from it.   

Another interest in resurgence is connection to nontrivial phenomena in which 
nonperturbative effect would play an important role such as confinement 
or symmetry breaking. It would be nice to clarify how resurgence are related 
to these interesting physics. 

From these motivations, in the previous papers, we considered a supersymmetric 
double-well matrix model \cite{Kuroki:2009yg,Kuroki:2010au} 
and derived one-point functions at arbitrary genus \cite{Kuroki:2016ucm}. 
Then we confirmed resurgence works for the one-point functions 
by calculating both their perturbative expansion 
and one-instanton contribution \cite{Kuroki:2019ets}. 
This model is proposed as nonperturbative formulation of 
type IIA superstring theory on a Ramond-Ramond background 
\cite{Endres:2013sda,Kuroki:2013qpa}. 
In fact, we have confirmed that the double scaling limit of this model 
can be taken and the perturbative expansion of the one-point functions has 
stringy behavior, namely expansion in terms of the string coupling constant 
with finite coefficient which grows as $(2h)!$ at sufficiently high genus $h$.  
Furthermore, the most important property of this model is that 
under the double scaling limit, the supersymmetry (SUSY) of this model is shown 
to be broken spontaneously and nonperturbatively
\cite{Endres:2013sda,Nishigaki:2014ija}. 
Thus our main interest in this model is relation 
between spontaneous SUSY breaking and resurgence structure. 
In fact, by computing an order parameter of SUSY breaking, 
we have recognized that an instanton in the matrix model triggers 
the SUSY breaking and that resurgence structure correctly reproduces 
its effect in the one-point functions. Since the instanton effect 
takes form of $\exp(-C/g_s)$ where $C$ is a positive constant 
and $g_s$ is the string coupling constant, 
the SUSY breaking would be possibly caused by generation or condensation 
of a D-brane-like object. Thus there is possibility that this model would 
provide new and interesting mechanism of spontaneous SUSY breaking 
in superstring theory. The key in applying resurgence to this model 
is to consider correlation functions of non-supersymmetric (non-SUSY) operators. 
In fact, although the SUSY is broken in our model, 
the breaking is nonperturbative and the SUSY is preserved 
at all order in perturbation theory \cite{Endres:2013sda}. 
Reflecting this fact, in the computation of the perturbative expansion 
of the correlation functions of non-SUSY operators, 
the Nicolai mapping is still available and calculation 
is reduced to that of the Gaussian matrix model in which nice results 
have been already obtained \cite{HT_2012}. 
We found that perturbative expansions of one-point functions of non-SUSY 
operators show non-Borel summable, stringy behavior and that resurgence works here. 
Thus we can overcome the problem mentioned above on non-Borel summable 
perturbative expansion in supersymmetric, non-trivial theory.\footnote
{One of other approaches to overcome this issue is to introduce 
a small parameter explicitly breaking SUSY
\cite{Fujimori:2016ljw,Dunne:2016jsr}. } 

In this paper we extend the results in \cite{Kuroki:2016ucm,Kuroki:2019ets} 
and study resurgence structure of two-point functions of non-SUSY operators 
in our model under the double scaling limit. More precisely, we compute the two-point 
functions both at arbitrary genus and on the one-instanton background. 
We then identify ambiguities in both results and confirm they cancel each other.  
There are several motivations of this extension. 
From string theory point of view, first of all, the result of multi-point functions 
at arbitrary genus itself is invaluable. Although our model has SUSY, 
it enjoys spontaneous SUSY breaking and its S-matrix is not trivial. 
It would be rare that we can get perturbative expansion at all order 
in such a nontrivial theory. 
Another motivation of considering the two-point function is the presence 
of logarithmic scaling violation. In \cite{Kuroki:2012nt} we showed that 
some planar two-point functions have new critical behavior 
as a power of log. Recalling the case of the two-dimensional 
bosonic string\cite{Gross:1990ub}, we may anticipate such logarithmic behavior 
would disappear at higher genus. 
{}From our result, we can explicitly check if this is the case.  

Our study also has motivations from resurgence. In general, resurgence structure 
becomes different for different quantities even in the same theory. 
Thus it would be intriguing to check how resurgence structure changes 
according to correlation functions in question. Furthermore, it should be 
confirmed that from perturbative expansions of several correlation functions, 
we can deduce the same nonperturbative effect if its origin is identical. 
We also expect that this kind of analysis would provide a clue to 
unified picture of resurgence structure for all correlation functions. 
Finally, by considering the two-point function, we have two integration variables 
and then we have more saddle points describing instantons, which leads to 
rich structure of resurgence. Hence this kind of study would give some insight 
for future study of resurgence.  

The organization of this paper is as follows. In the next section, we give a brief review 
of the supersymmetric double-well matrix model. Correlation functions are expressed 
in terms of eigenvalues and are defined in each instanton sector. 
In section \ref{sec:Nicolai}, we explain how to compute correlation functions 
in our matrix model by utilizing the Nicolai mapping. 
In section \ref{sec:zero-inst}, we consider contribution from the zero-instanton sector 
to the two-point function, and find that there exists an ambiguity 
after applying the Borel resummation technique. 
Then in section \ref{sec:one-inst}, we see that contribution 
from the one-instanton sector also has another ambiguity, 
and confirm that these ambiguities exactly cancel each other at the leading order. 
The last section is devoted to conclusions and discussions.

%%%%%%%%%%%%%%%%%%%%%%%%%%%%%%%%%%%%%%%%%%%%%%%%%%%%%%%%%%%%%%%%%%%%%%%%%%%%%
\section{Review of the supersymmetric matrix model}
\label{sec:MM_review}
\setcounter{equation}{0}
%%%%%%%%%%%%%%%%%%%%%%%%%%%%%%%%%%%%%%%%%%%%%%%%%%%%%%%%%%%%%%%%%%%%%%%%%%%%%
In this section, we give a brief review of the supersymmetric double-well 
matrix model which has been proposed as a nonperturbative formulation 
of type IIA superstring theory in two dimensions. Definitions and notations are 
exactly the same as in \cite{Kuroki:2019ets}. Most of the review in this section 
overlaps with section 2 there. 

%%%%%%%%%%%%%%%%%%%%%%%%%%%%%%%%%%%%%%%%%
\subsection{The model and double scaling limit}
We study a supersymmetric double-well matrix model with the action 
\begin{align}
S = N \text{tr}\left[\frac12 B^2 +iB(\phi^2-\mu^2) +\bar\psi (\phi\psi+\psi\phi)\right], 
\label{eq:S}
\end{align}
where $B$ and $\phi$ are $N\times N$ Hermitian matrices, 
and $\psi$ and $\bar\psi$ are $N\times N$ Grassmann-odd matrices. 
$\mu^2$ is a parameter of the model. 
The action $S$ is invariant under SUSY transformations generated by $Q$ and $\bar{Q}$: 
\begin{align}
&Q\phi =\psi, &&Q\psi=0, &&Q\bar{\psi} =-iB, &&QB=0, \nn \\
&\bar{Q} \phi = -\bar{\psi}, &&\bar{Q}\bar{\psi} = 0, 
&&\bar{Q} \psi = -iB, &&\bar{Q} B = 0,  
\label{eq:SUSY}
\end{align}
which are nilpotent: $Q^2=\bar{Q}^2=\{ Q,\bar{Q}\}=0$. 
After integrating out the auxiliary variable $B$ in (\ref{eq:S}), 
the scalar potential of $\phi$ reads
\begin{align}
V(\phi)=\frac12(\phi^2-\mu^2)^2. 
\label{eq:DW}
\end{align}
In the planar limit ($N\rightarrow\infty$ with $\mu^2$ fixed) 
of this model, there are infinitely degenerate supersymmetric vacua 
parametrized by filling fractions $(\nu_+, \nu_-)$ for $\mu^2\geq 2$. 
They represent configurations that $\nu_\pm N$ 
of the eigenvalues of $\phi$ are around the minimum 
$\pm |\mu |$ of the double-well potential \eqref{eq:DW} 
\cite{Kuroki:2009yg, Kuroki:2010au}. 
On the other hand, for $\mu^2<2$ we have a unique vacuum without SUSY. 
The boundary $\mu^2=2$ is a critical point of the third-order phase transition. 
In the planar limit, it is explicitly shown in \cite{Kuroki:2013qpa,Kuroki:2012nt} that 
several types of correlation functions in the matrix model reproduce 
tree amplitudes in two-dimensional type IIA superstring theory 
on a nontrivial Ramond-Ramond background.  
In addition, we have considered the following double scaling limit
\cite{Endres:2013sda} that approaches the critical point 
from the inside of the supersymmetric phase: 
\begin{align}
N\rightarrow\infty, \quad \mu^2\rightarrow 2+0, \quad  \mbox{with} \quad  
s=N^{\frac23}(\mu^2-2):\mbox{~fixed}.
\label{eq:dsl}
\end{align}
In this limit the matrix model is expected to provide a nonperturbative formulation 
of the superstring theory with string coupling constant $g_s$ proportional to $s^{-\frac32}$, 
where the two supersymmetries in \eqref{eq:SUSY} correspond to the target-space 
supersymmetries\cite{Kuroki:2013qpa,Kuroki:2012nt}. 
From this viewpoint the planar limit mentioned above is regarded as 
$g_s\rightarrow 0$ limit. In fact, in \cite{Kuroki:2016ucm} one-point functions 
for the single-trace operators of powers of $\phi$ are explicitly calculated 
at arbitrary genus and are found to be finite at each genus 
under the double scaling limit \eqref{eq:dsl}. 
Furthermore, its coefficients show stringy behavior $(2h)!$ with genus $h\gg 1$. 
In\cite{Endres:2013sda,Nishigaki:2014ija}, we also found that contribution 
from matrix-model instantons (isolated eigenvalues of $\phi$ located 
at the top of the effective potential) to the free energy is finite and takes form of  
$\exp\left(-C/g_s\right)$ with a positive constant $C$ of $\cO(1)$. 
The most remarkable feature of the model ever found is that these instantons cause 
spontaneous supersymmetry breaking in the matrix model, 
which implies violation of target-space supersymmetry 
induced by D-brane like objects in the corresponding superstring theory.\footnote
{The correspondence between isolated eigenvalues and solitons or D-branes is 
well-established in bosonic noncritical string theories 
\cite{David:1992za,Kazakov:2004du,Hanada:2004im,Kawai:2004pj,Sato:2004tz,Ishibashi:2005dh,Ishibashi:2005zf,Kuroki:2006wn}. }

%%%%%%%%%%%%%%%%%%%%%%%%%%%%%%%%%%%%%
\subsection{Correlation functions in fixed filling fraction}
In this subsection, we define correlation functions of our model \eqref{eq:S} 
in a fixed filling fraction. 
The partition function is expressed in terms of the eigenvalues of $\phi$ as 
\begin{align}
Z&\equiv (-1)^{N^2}\int d^{N^2}B\,d^{N^2}\phi\,
\left(d^{N^2}\psi\,d^{N^2}\bar{\psi}\right)\,e^{-S} \nn \\
&=\tilde C_N\int_{-\infty}^{\infty}
\left(\prod_{i=1}^N2\lambda_id\lambda_i\right)\triangle(\lambda^2)^2\,
e^{-N\sum_{i=1}^N\frac12(\lambda_i^2-\mu^2)^2}, 
\label{eq:Z}
\end{align}
where the normalization of the integration measure is fixed as 
\begin{align}
&\int d^{N^2}\phi\,e^{-N\tr\left(\frac12\phi^2\right)}
= \int d^{N^2}B\,e^{-N\tr\left(\frac12 B^2\right)}=1 , \nn \\
&(-1)^{N^2}\int \left(d^{N^2}\psi\,d^{N^2}\bar{\psi}\right)\,
e^{-N\tr \left(\bar{\psi}\psi\right)}=1.
\end{align}
$\tilde C_N$ is a constant dependent only on $N$: 
$\tilde{C}_N=(2\pi)^{-\frac{N}{2}}N^{\frac{N^2}{2}}\left(\prod_{k=0}^Nk!\right)^{-1}$
\cite{Kuroki:2010au}, and 
$\triangle(x)$ is the Vandermonde determinant for eigenvalues $x_i$ ($i=1,\cdots, N$): 
$\triangle(x) \equiv \prod_{i>j}(x_i-x_j)$. 
By dividing the integration region of each $\lambda_i$ according to the filling fraction, 
the total partition function can be expressed as a sum of 
each partition function with a fixed filling fraction: 
\begin{align}
&Z=\sum_{\nu_-N=0}^{N}\frac{N!}{(\nu_+N)!(\nu_-N)!}\,Z_{(\nu_+,\nu_-)}, \nn \\ 
&Z_{(\nu_+,\nu_-)}\equiv \tilde C_N
\int_0^{\infty}\left(\prod_{i=1}^{\nu_+N}2\lambda_id\lambda_i\right)
\int_{-\infty}^0\left(\prod_{j=\nu_+N+1}^N2\lambda_jd\lambda_j\right)
\triangle(\lambda^2)^2 
\,e^{-N\sum_{m=1}^N\frac12(\lambda_m^2-\mu^2)^2}. 
\label{eq:ZinFFsector}
\end{align}
By changing the integration variables $\lambda_j\rightarrow -\lambda_j$ 
($j=\nu_+N+1,\cdots, N$), it is easy to find that 
$Z_{(\nu_+, \nu_-)}=(-1)^{\nu_-N}Z_{(1,0)}$
and that the total partition function vanishes. 

We then define the correlation function of $K$ single-trace operators for functions 
of $\phi$: $\frac{1}{N}\tr f_a(\phi)$ ($a=1,\cdots, K$) in the filling fraction 
$(\nu_+,\nu_-)$ as 
\begin{align}
\vev{\prod_{a=1}^K\frac{1}{N}\tr f_a(\phi)}^{(\nu_+,\nu_-)} 
\equiv& \frac{\tilde C_N}{Z_{(\nu_+,\nu_-)}}
\int_0^{\infty}\left(\prod_{i=1}^{\nu_+N}2\lambda_id\lambda_i\right)
\int_{-\infty}^0\left(\prod_{j=\nu_+N+1}^N2\lambda_jd\lambda_j\right)
\triangle(\lambda^2)^2\nn \\
&\times \left(\prod_{a=1}^K\frac{1}{N}\sum_{i=1}^Nf_a(\lambda_i)\right)\,
e^{-N\sum_{m=1}^N\frac12(\lambda_m^2-\mu^2)^2},  
\label{eq:correlator}
\end{align}
and extract its connected part in the $1/N$-expansion:  
\begin{align}
\vev{\prod_{a=1}^K \frac{1}{N}\tr f_a(\phi)}_c^{(\nu_+,\nu_-)}
=\sum_{h=0}^\infty \frac{1}{N^{2h+2K-2}}\,
\vev{\prod_{a=1}^K \frac{1}{N}\tr f_a(\phi)}_{c,h}^{(\nu_+,\nu_-)},  
\label{eq:vev_MM}
\end{align}
where $\vev{\,\cdot\,}_{c,\,h}^{(\nu_+,\nu_-)}$ denotes the connected correlation function 
on a handle-$h$ random surface with the $N$-dependence factored out; 
i.e., the quantity of ${\cal O}(N^0)$. 
Hereafter we consider the case where $f_a(\phi)$ is a monomial of $\phi$: 
$f_a(\phi)=\phi^p$ ($p\in\bm N$). When $p$ is even, the correlation functions 
of $\frac{1}{N}\tr\phi^{p}$'s are expressed as linear combinations of those of operators 
$\frac{1}{N}\tr B^k$ ($k\in\bm N$), which are supersymmetric. 
Hence they become just polynomials of $s$.  
On the other hand, when $p$ is odd,  $\frac{1}{N}\tr\phi^{p}$ is not supersymmetric, 
and their correlation functions show nonanalytic behavior on $s$~\cite{Kuroki:2012nt}. 
Thus in this paper in order to study resurgence structure, 
we focus on the two-point function of the odd-power operators 
$\vev{\frac1N\tr\phi^p\frac1N\tr\phi^q}_{c,h}^{(1,0)}$ for odd $p,q$: 
\eqref{eq:vev_MM} with $K=2$, $f_1(\phi)=\phi^{p}$, and $f_2(\phi)=\phi^q$ 
in the filling fraction $(\nu_+,\nu_-)=(1,0)$.\footnote
{It is shown in \cite{Kuroki:2012nt} that at least at the planar level ($h=0$) 
and up to the three-point functions 
($1\leq K\leq 3$), it is easy to recover filling fraction dependence of correlation functions 
from those in $(\nu_+,\nu_-)=(1,0)$.}

%%%%%%%%%%%%%%%%%%%%%%%%%%%%%%%%%%%%%%%%%
\subsection{Correlation functions in fixed instanton sector}
In this subsection, we divide correlation functions in the $(\nu_+,\nu_-)=(1,0)$ sector 
into contributions from sectors with definite instanton numbers as done 
in \cite{Endres:2013sda}. 
In \eqref{eq:ZinFFsector}, the partition function $Z_{(1,0)}$ is expressed 
as the integrations of $N$ eigenvalues along the positive real axis $\bm R_+=[0,\infty)$. 
The eigenvalue distribution in the planar limit is given as \cite{Kuroki:2009yg,Kuroki:2012nt} 
\begin{align}
\left.\vev{\frac{1}{N}\sum_{i=1}^N\delta(x-\lambda_i)}^{(1,0)}\right|_{\text{planar}}=
\begin{cases}
\frac{x}{\pi}\sqrt{(x^2-a^2)(b^2-x^2)} & (a\leq x\leq b) \\
0 & (\text{otherwise})
\end{cases},
\label{eq:rho}
\end{align}
with $a=\sqrt{\mu^2-2}$ and $b=\sqrt{\mu^2+2}$, which means that 
all the eigenvalues are confined in the interval $[a,b]$. Dividing 
the integration region of each eigenvalue $\bm R_+$ into the inside and outside 
of the interval:
\begin{align}
\int_0^\infty d\lambda_i=\int_a^b d\lambda_i+\int_{\bm R_+\setminus [a,b]} d\lambda_i, 
\end{align}
we decompose the partition function as 
\begin{align}
&Z_{(1,0)}=\sum_{p=0}^N\left.Z_{(1,0)}\right|_{p\text{-inst.}}, \label{eq:instantonsum} \\
&\left.Z_{(1,0)}\right|_{p\text{-inst.}}={}
\begin{pmatrix}N \\ p\end{pmatrix}\tilde C_N
\int_a^b\prod_{i=1}^{N-p}2\lambda_id\lambda_i
\int_{\bm R_+\setminus [a,b]}\prod_{j=1}^{p}2\lambda_jd\lambda_j\,
\Delta(\lambda^2)^2 
e^{-N\sum_{m=1}^N\frac12(\lambda_m^2-\mu^2)^2}.
\label{eq:instantonsectors}
\end{align}
Each contribution with fixed $p$ is regarded as the partition function 
in the $p$-instanton sector.  
In fact, an instanton in our model corresponds to a saddle point of effective potential 
$V_{\text{eff}}(\lambda_i)$ with respect to a single eigenvalue $\lambda_i$, 
which is obtained by integrating out all the eigenvalues other than $\lambda_i$ in \eqref{eq:Z}. 
Its saddle point turns out to be the origin $\lambda_i=0$ \cite{Endres:2013sda}. 
For large $s$ (small $g_s$) under the double scaling limit \eqref{eq:dsl}, 
the main contribution from the outside of the interval $\bm R_+\setminus [a,b]$ 
is provided by such an instanton located at the origin.  
According to \cite{Endres:2013sda,Nishigaki:2014ija}, 
the partition function \eqref{eq:instantonsectors} in the $p$-instanton sector reads 
\begin{align}
\left.Z_{(1,0)}\right|_{\text{$p$-inst.}}
=\left(\frac{e^{-\frac43s^{\frac32}}}{16\pi s^{\frac32}}\right)^p
\left(1+\cO\left(s^{-\frac32}\right)\right) 
\label{eq:Z_pinst_dsl}
\end{align}
in the double scaling limit with $s\gg 1$ fixed. 
Hence \eqref{eq:instantonsum} is a trans-series expanded by the instanton weight 
$e^{-\frac43s^{\frac32}}/(16\pi s^{\frac32})$. Similarly, the correlation functions 
are also decomposed by contribution from each instanton sector, 
which is written as  a trans-series:
\begin{align}
&\vev{\prod_{a=1}^K{\cO}_a}^{(1,0)}_c
=\sum_{p=0}^N\left.\vev{\prod_{a=1}^K{\cO}_a}^{(1,0)}_c\right|_{\text{$p$-inst.}}, 
\label{eq:O_instantonsum} \\
&\left.\vev{\prod_{a=1}^K{\cO}_a}^{(1,0)}_c\right|_{\text{$p$-inst.}} 
\propto \left(e^{-\frac43s^{\frac32}}\right)^p\quad 
\text{under the double scaling limit}. 
\label{eq:O_exp_inst}
\end{align} 
In \eqref{eq:O_exp_inst} we note that the power of $s$ other than the instanton factor 
$\left(e^{-\frac43s^{\frac32}}\right)^p$ depends on ${\cO}_a$'s 
and is in general different from that of the partition function in \eqref{eq:Z_pinst_dsl}.  
In this paper we argue that in the two-point functions of the odd operators 
${\cO}_1=\frac1N\tr\phi^p$, ${\cO}_2=\frac1N\tr\phi^q$ ($p,q$: odd), 
both $\left.\vev{\prod_{a=1}^2{\cO}_a}^{(1,0)}_c\right|_{\text{$0$-inst.}}$ 
and $\left.\vev{\prod_{a=1}^2{\cO}_a}^{(1,0)}_c\right|_{\text{$1$-inst.}}$ have ambiguities, 
but they cancel each other and hence the trans-series in \eqref{eq:O_instantonsum} 
is well-defined at the leading order of the large-$s$ expansion 
up to the one-instanton contribution $p=0,1$.

%%%%%%%%%%%%%%%%%%%%%%%%%%%%%%%%%%%%%%%%%%%%%%%%%%%%%%%%%%%%%%%%%%%%%%%%%%%%%
\section{Non-SUSY correlation functions via the Gaussian matrix model}
\label{sec:Nicolai}
\setcounter{equation}{0}
%%%%%%%%%%%%%%%%%%%%%%%%%%%%%%%%%%%%%%%%%%%%%%%%%%%%%%%%%%%%%%%%%%%%%%%%%%%%%
In this section, we show that correlation functions of the operators 
$\frac1N\tr\phi^p$ with odd $p$ in the filling fraction $(\nu_+,\nu_-)=(1,0)$ 
can be expressed by those of the Gaussian matrix model. 
The point is that the Nicolai mapping can be applied there 
even if the operators are not supersymmetric.  

%%%%%%%%%%%%%%%%%%%%%%%%%%%%%%%%%%%%%
\subsection{$\phi^2$-resolvent}
Extending the derivation in \cite{Kuroki:2016ucm,Kuroki:2019ets}, 
we begin with the connected $K$-point function of the $\phi^2$-resolvent 
in the filling fraction $(1,0)$ 
\begin{align}
\vev{\prod_{a=1}^KR_2(z_a^2)}^{(1,0)}_c
=\vev{\prod_{a=1}^K\left(\frac1N\tr\frac{1}{z_a^2-\phi^2}\right)}^{(1,0)}_c. 
\label{eq:R2}
\end{align}
In terms of the eigenvalues, $R_2(z^2)$ becomes  
\begin{align}
\frac1N\sum_{i=1}^N \frac{1}{z^2-\lambda_i^2}
= \frac1N\frac{1}{2z}\sum_{i=1}^N\left(\frac{1}{z-\lambda_i}+\frac{1}{z+\lambda_i}\right), 
\end{align}
and $1/\left(z-\lambda_i\right)$ ($1/\left(z+\lambda_i\right)$) has poles only 
on the positive (negative) real axis for the filling fraction $(1,0)$. 
Thus suppose $C_0$ is a contour which encloses only the poles 
at $z=\lambda_i$ for ${}^\forall i$ counterclockwise and $f(z)$ is an any function of $z$, 
\begin{align}
&\prod_{a=1}^K\left(\frac{1}{2\pi i}\oint_{C_0}dz_a\,2z_af_a(z_a)\right)
\vev{\prod_{a=1}^KR_2(z_a^2)}_c^{(1,0)} \nn \\ 
&=\vev{\prod_{a=1}^K\left(\frac{1}{2\pi i}\oint_{C_0}dz_a\,f_a(z_a)
\frac1N\sum_{i_a=1}^N\frac{1}{z_a-\lambda_{i_a}}\right)}_c^{(1,0)} \nn \\
&=\vev{\prod_{a=1}^K\frac1N\sum_{i_a=1}^Nf_a(\lambda_{i_a})}_c^{(1,0)}
=\vev{\prod_{a=1}^K\frac1N\tr f_a(\phi)}_c^{(1,0)}. 
\label{eq:byresolvent}
\end{align}

\begin{figure}[h]
\centering
\includegraphics[width=10cm, bb=200 200 1000 600, clip]{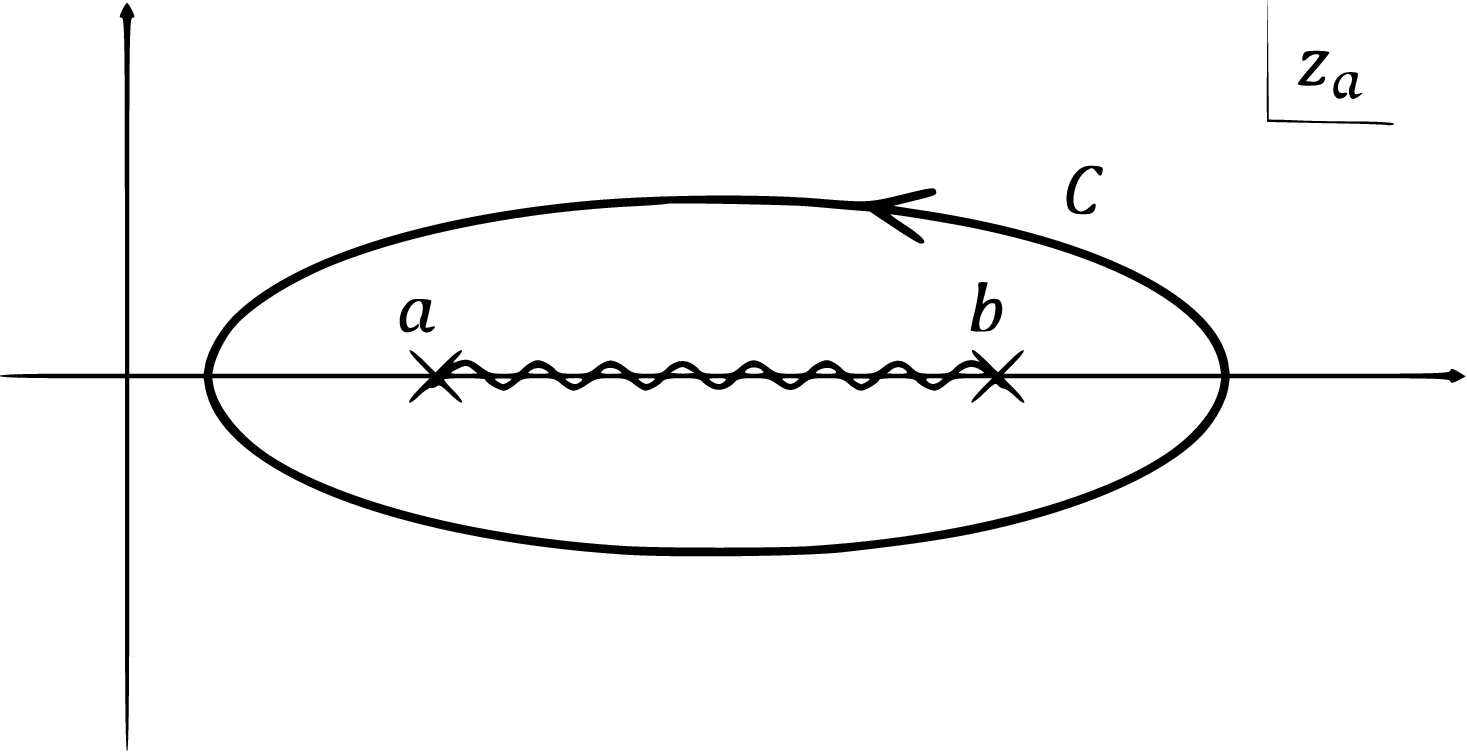}
%\vspace{-4cm}
\caption{\small Integration contour $C$ on the complex $z_a$-plane.}
\label{fig:zcontour}
\end{figure}

In particular, in the zero-instanton sector, all $\lambda_i$ are in the interval $[a,b]$ 
mentioned below \eqref{eq:rho}, and therefore, 
\begin{align}
\left.\vev{\prod_{a=1}^K\frac1N\tr f_a(\phi)}_c^{(1,0)}\right|_{\text{0-inst.}}
=\prod_{a=1}^K\left(\frac{1}{2\pi i}\oint_{C}dz_a\,2z_af_a(z_a)\right)
\vev{\prod_{a=1}^KR_2(z_a^2)}_c^{(1,0)}
\label{eq:by0instresolvent}
\end{align}
where $C$ denotes a contour encircling the interval $[a,b]$ counterclockwise 
as depicted in Fig.~\ref{fig:zcontour}. 
Since $C$ does not contain $z_a=0$ inside, contribution from the instanton at the origin 
is not included in \eqref{eq:by0instresolvent}. 
%%%%%%%%%%%%%%%%%%%%%%%%%%%%%%%%%%%%%%%%%%%%%%%%%%%%%%%%%%%%%%%%%%%%%%%
%%%%%%%%%%%%%%%%%%%%%%%%%%%%%%%%%%%%%%%%%%%%%%%%%%%%%%%%%%%%%%%%%%%%%%%%
The key here is that even if $\frac1N\tr f_a(\phi)$'s are non-SUSY, 
\eqref{eq:by0instresolvent} enables us to compute their correlation functions 
via those of the $\phi^2$-resolvent, and that the latter can be expressed 
by the Gaussian matrix model via the Nicolai mapping.  

%%%%%%%%%%%%%%%%%%%%%%%%%%%%%%%%%%%%%
\subsection{Nicolai mapping}
\begin{figure}[h]
\centering
%\vspace{-0.5cm}
\includegraphics[width=10cm, bb=200 200 1000 600, clip]{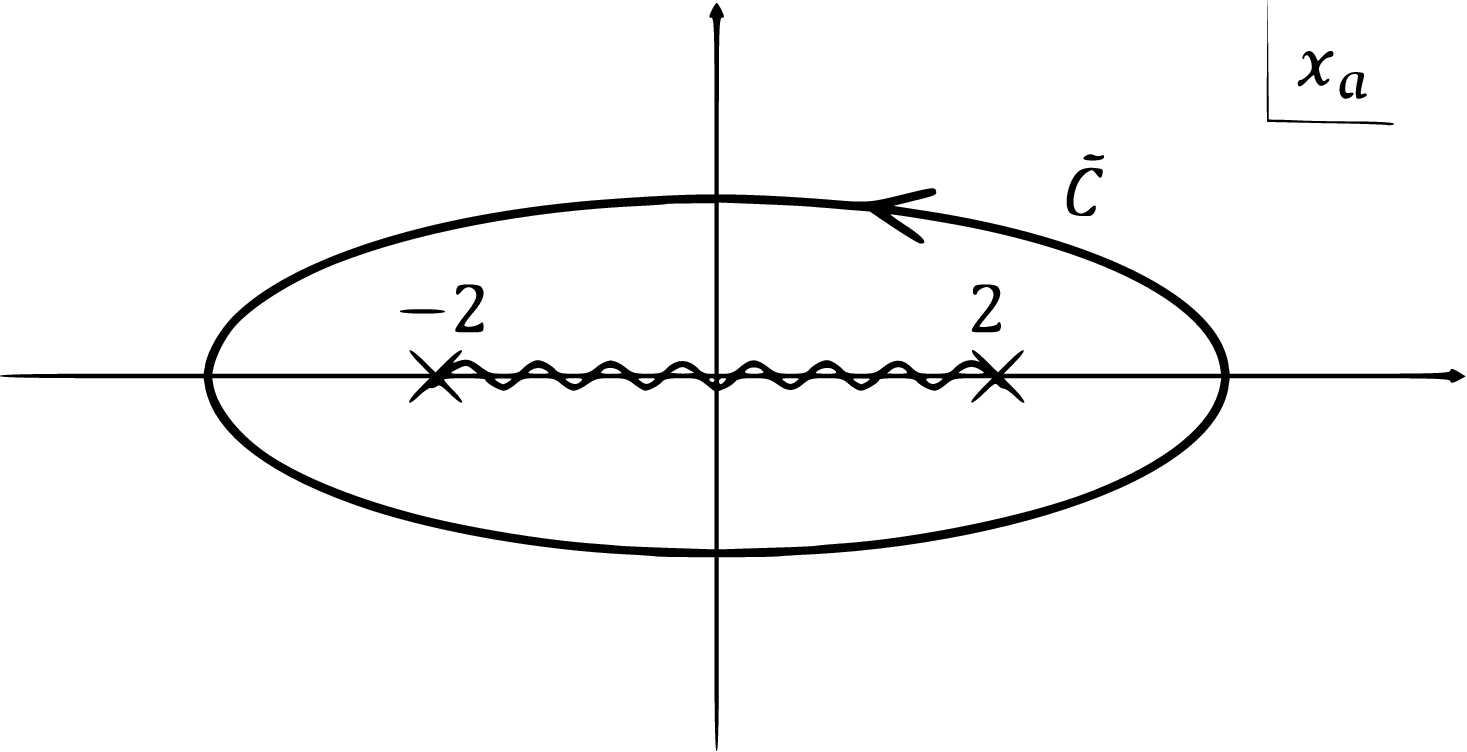}
%\vspace{-4cm}
\caption{\small Integration contour $\tilde C$ on the complex $x_a$-plane.}
\label{fig:xcontour}
\end{figure}

By setting $x_i=\mu^2-\lambda_i^2$, we have 
\begin{align}
&Z^{(1,0)}\vev{\prod_{a=1}^KR_2(z_a^2)}^{(1,0)} \nn \\
&=\tilde C_N\int_0^{\infty}\left(\prod_i2\lambda_id\lambda_i\right)
\Delta(\lambda^2)^2
\left(\prod_{a=1}^K\frac1N\sum_{i_a=1}^N\frac{1}{z_a^2-\lambda_{i_a}^2}\right)
e^{-N\sum_i\frac12(\lambda_i^2-\mu^2)^2} \nn \\
&=\tilde C_N\int_{-\infty}^{\mu^2}\prod_idx_i\,\Delta(x)^2
\left(\prod_{a=1}^K\frac1N\sum_{i_a=1}^N\frac{1}{z_a^2-\mu^2+x_{i_a}}\right)
e^{-N\sum_i\frac12x_i^2} \nn \\
&=(-1)^KZ^{(G')}\vev{\prod_{a=1}^KR_M(\mu^2-z_a^2)}^{(G')}.
\label{eq:resolventbyNicolai} 
\end{align}
Here $(G')$ denotes quantities of the Gaussian matrix model 
of an $N\times N$ Hermitian matrix $M$ with the upper boundary $\mu^2$ 
for its eigenvalues $x_i$ ($i=1,\cdots,N$): 
\begin{align}
&Z^{(G')}=\tilde C_N\int_{-\infty}^{\mu^2}\prod_idx_i\,
\Delta(x)^2\,e^{-N\sum_i\frac12x_i^2} \nn \\
&\vev{\prod_{a=1}^K\frac1N\tr f_a(M)}^{(G')}=\frac{1}{Z^{(G')}}
\int_{-\infty}^{\mu^2}\prod_idx_i\,
\Delta(x)^2\prod_{a=1}^K\left(\frac1N\sum_if_a(x_i)\right)e^{-N\sum_i\frac12x_i^2},
\label{eq:modifiedGaussian}
\end{align} 
and $R_M(z)$ is the resolvent of this Gaussian matrix model: 
$R_M(z)\equiv\frac1N\tr\frac{1}{z-M}$. 
By considering $K=0$ case above, we find $Z^{(1,0)}=Z^{(G')}$ and hence 
\begin{align}
\vev{\prod_{a=1}^KR_2(z_a^2)}^{(1,0)}
=(-1)^K\vev{\prod_{a=1}^KR_M(\mu^2-z_a^2)}^{(G')}.
\end{align}
Taking the connected part, we find 
\begin{align}
\vev{\prod_{a=1}^KR_2(z_a^2)}^{(1,0)}_c
=(-1)^K\vev{\prod_{a=1}^KR_M(\mu^2-z_a^2)}^{(G')}_c.
\label{eq:resolventrelation}
\end{align}
Plugging this equation into \eqref{eq:by0instresolvent}, we obtain  
\begin{align}
\left.\vev{\prod_{a=1}^K\frac1N\tr f_a(\phi)}_c^{(1,0)}\right|_{\text{0-inst.}}
=\prod_{a=1}^K\left(-\frac{1}{2\pi i}\oint_{C}dz_a\,2z_af_a(z_a)\right)
\vev{\prod_{a=1}^KR_M(\mu^2-z_a^2)}_c^{(G)}.
\label{eq:formula1}
\end{align}
Here as observed in \cite{Endres:2013sda}, in the Gaussian matrix model 
\eqref{eq:modifiedGaussian}, the boundary at $x_i=\mu^2$ yields 
nonperturbative effect with respect to the $1/N$-expansion and, therefore, we can  
replace $\mu^2$ with $+\infty$ as far as the $1/N$-expansion is concerned. 
Thus on the right-hand side in \eqref{eq:formula1} 
we consider the standard Gaussian matrix model 
where $x_i$ runs from $-\infty$ to $\infty$ and this is the reason 
why we put $(G)$ instead of $(G')$ on the correlation function.\footnote
{In the case of the one-point functions, this is justified explicitly by using 
the orthogonal polynomials up to the one-instanton sector in section 3 
in \cite{Kuroki:2019ets}.} 
If we further change the integration variables as $x_a=\mu^2-z_a^2$, 
eq.\eqref{eq:formula1} finally becomes 
\begin{align}
\left.\vev{\prod_{a=1}^K\frac1N\tr f_a(\phi)}_c^{(1,0)}\right|_{\text{0-inst.}}
=\prod_{a=1}^K\left(-\frac{1}{2\pi i}\oint_{\tilde C}dx_a\,
f_a((\mu^2-x_a)^\frac{1}{2})\right)\vev{\prod_{a=1}^KR_M(x_a)}_c^{(G)},
\label{eq:formula2}
\end{align}
where $\tilde C$ is a contour on the complex $x_a$-plane shown in Fig.~\ref{fig:xcontour}.  
%%%%%%%%%%%%%%%%%%%%%%%%%%%%%%%%%%%%%%%%%%%%%%%%%%%%%%%%%%%%%%%%%%%%%%%
%%%%%%%%%%%%%%%%%%%%%%%%%%%%%%%%%%%%%%%%%%%%%%%%%%%%%%%%%%%%%%%%%%%%%%%%
{}From this equation, once we get perturbative expansion 
of the correlation function of the resolvents in the Gaussian matrix model, 
we can derive perturbative series of the correlation function 
of any operators of $\phi$ in the zero-instanton sector in our model, 
if we can perform the $x_a$-integrations.  

%%%%%%%%%%%%%%%%%%%%%%%%%%%%%%%%%%%%%%%%%%%%%%%%%%%%%%%%%%%%%%%%%%%%%%%%%%%%%
\section{Two-point function in the zero-instanton sector}
\label{sec:zero-inst}
\setcounter{equation}{0}
%%%%%%%%%%%%%%%%%%%%%%%%%%%%%%%%%%%%%%%%%%%%%%%%%%%%%%%%%%%%%%%%%%%%%%%%%%%%%
In \cite{Kuroki:2016ucm}, based on \eqref{eq:formula2} we have obtained 
the genus expansion at all order of the one-point function of the odd operator 
$\frac1N\tr\phi^p$ ($p$: odd) in the zero-instanton sector 
under the double scaling limit (\ref{eq:dsl}). 
In this section we extend this result and apply \eqref{eq:formula2} to the two-point function 
of the odd operators to find its perturbative series. 
The genus zero results have been already presented in \cite{Kuroki:2012nt}. 

%%%%%%%%%%%%%%%%%%%%%%%%%%%%%%%%%%%%%%%%%%%%%%%%%%%%%%%%%%%%%%%%%%%%%%
\subsection{One-point function at arbitrary genus}
Before discussing the two-point function, we briefly review the result 
of the one-point function given in \cite{Kuroki:2016ucm}. 
Since the genus expansion for the one-point function of the resolvent 
in the standard Gaussian matrix model $\vev{R_M(z)}^{(\text{G})}_c$ has been obtained 
at arbitrary genus in the literature e.g. \cite{HT_2012}, 
we applied it to \eqref{eq:formula2} and arrived at the expression 
\begin{align}
&N^{\frac23(k+2)}\left.\vev{\frac1N\tr\phi^{2k+1}}^{(1,0)}\right|_{\text{0-inst., univ.}} \nn \\
&=\frac{\Gamma\left(k+\frac32\right)}{2\pi^{\frac32}} 
\Biggl\{\sum_{h=0}^{\left[\frac13(k+2)\right]}\frac{1}{h!}\left(-\frac{1}{12}\right)^h
\frac{1}{\Gamma\left(k+3-3h\right)}s^{k+2-3h}\ln s \nn \\
&\phantom{N^{-\frac23(k+2)}} \hspace{10mm} 
+(-1)^{k+1}\sum_{h=\left[\frac13(k+2)\right]+1}^{\infty}\frac{1}{h!}\left(\frac{1}{12}\right)^h
\Gamma\left(3h-k-2\right)s^{k+2-3h}\Biggr\}
\label{eq:1ptgenusexp}
\end{align}
in the double scaling limit~(\ref{eq:dsl}), 
where $[x]$ denotes the greatest integer less than or equal to $x$. 
We see that the infinite series in the curly brackets on the right-hand side gives 
the genus expansion where the power of $g_s^2\propto s^{-3}$ counts 
the number of handles. 
The suffix ``univ.'' on the left-hand side means that we take the most dominant 
nonanalytic term at $s=0$ in the limit \eqref{eq:dsl} (the universal part). 
Note that in order to get a finite result, we also need the overall factor $N^{\frac23(k+2)}$, 
which is interpreted as ``wave function renormalization'' of the operator 
$\frac1N\tr\phi^{2k+1}$ itself.

%%%%%%%%%%%%%%%%%%%%%%%%%%%%%%%%%%%%%
\subsection{Two-point function at arbitrary genus}
By taking the genus expansion in both sides, \eqref{eq:formula2} leads to 
\begin{align}
&\left.\vev{\frac1N\tr f(\phi)\frac1N\tr g(\phi)}_{c,h}^{(1,0)}\right|_{\text{0-inst.}} \nn \\
&=\frac{1}{(2\pi i)^2}\oint_{\tilde C_x}dx\oint_{\tilde C_y}dy\,
f((\mu^2-x)^{\frac12})g((\mu^2-y)^{\frac12})
\vev{R_M(x)R_M(y)}_{c,h}^{(G)}, 
\label{eq:two-ptintegral}
\end{align} 
where $\tilde C_x$ and $\tilde C_y$ are the contours on the complex $x$- and $y$-plane 
as depicted in Fig.~\ref{fig:xcontour}, respectively. 
The two-point function of the resolvent in the Gaussian matrix model is given 
in \cite{HT_2012} as 
\begin{align}
&\vev{R_M(x)R_M(y)}_{c,h}^{(G)}=\frac{1}{2(x-y)^2}\Gamma_h(x,y) 
\label{eq:two-ptandGamma} \\
&\text{with} \quad \Gamma_0(x,y)=(2\eta'_0(x)-1)(2\eta_0'(y)-1)
-\widetilde\eta_0(x)\widetilde\eta_0(y)-1, 
\label{eq:Gamma0} \\
&\phantom{\text{with}} \quad \Gamma_h(x,y)
=2\eta'_h(x)(2\eta'_0(y)-1)+2\eta'_h(y)(2\eta'_0(x)-1)
+4\sum_{j=1}^{h-1}\eta'_j(x)\eta'_{h-j}(y) \nn \\
&\phantom{\text{with} \quad \Gamma_h(x,y)=}
+\sum_{j=0}^{h-1}\eta^{\prime\prime}_j(x)\eta^{\prime\prime}_{h-1-j}(y)
-\sum_{j=0}^{h}\widetilde\eta_j(x)\widetilde\eta_{h-j}(y) \qquad (h\in\bm N),
\label{eq:Gammahtemp} \\
&\phantom{\text{with}} \quad \eta_0(x)=\frac12x-\frac12(x^2-4)^{\frac12}, 
\qquad \eta_j(x)=\sum_{r=2j}^{3j-1}C_{j,r}(x^2-4)^{-r-\frac12}
\qquad (j\in\bm N), 
\label{eq:etah} \\
&\phantom{\text{with}} \quad \widetilde\eta_j(x)=\eta_j(x)-x\eta'_j(x) 
\qquad (j\in\bm Z_{\geq 0}),
\end{align}
where $C_{j,r}$ satisfies a recursion relation 
\begin{align}
C_{j+1,r}=\frac{(2r-3)(2r-1)}{r+1}\left((r-1)C_{j,r-2}+(4r-10)C_{j,r-3}\right) \quad 
(2j+2\leq r\leq 3j+2)
\label{eq:recursion}
\end{align}
with $C_{j,2j-1}=C_{j,3j}=0$ understood and the initial conditions are given by 
\begin{align}
C_{0,-1}=-\frac12, \qquad C_{1,2}=1.
\label{eq:Cinitial_cond}
\end{align} 
{}From \eqref{eq:etah}, we see that 
$2\eta'_0(x)-1$ can be included in $2\eta'_j(x)$ as $j=0$ case 
by recogizing that when $j=0$, the sum over $r$ becomes setting $r=-1$ 
in \eqref{eq:etah}. 
According to this convention, in \eqref{eq:Gammahtemp} 
the first and the second term on the right-hand side can be identified 
with the $j=h$ and $j=0$ case of the third term, and hence  
\begin{align}
\Gamma_h(x,y)=4\sum_{j=0}^{h}\eta'_j(x)\eta'_{h-j}(y) 
+\sum_{j=0}^{h-1}\eta^{\prime\prime}_j(x)\eta^{\prime\prime}_{h-1-j}(y)
-\sum_{j=0}^{h}\widetilde\eta_j(x)\widetilde\eta_{h-j}(y) \quad (h\in\bm N).
\label{eq:Gammah}
\end{align}
Plugging the explicit form of $\eta_j$ given in \eqref{eq:etah} 
into \eqref{eq:Gammah}, we obtain for $h\in\bm Z_{\geq 0}$
\begin{align}
\Gamma_h(x,y)=&4(x^2-4)^{-3h-\frac12}(y^2-4)^{-3h-\frac12} \nn \\
&\times\biggl[\sum_{j=0}^h\sum_{r=2j}^{3j-1}\sum_{t=2(h-j)}^{3(h-j)-1}C_{j,r}C_{h-j,t}
(x^2-4)^{3h-1-r}(y^2-4)^{3h-1-t} \nn \\
&\phantom{\times\biggl[}\times\Bigl\{
(2r+1)(2t+1)xy-\left((r+1)x^2-2\right)\left((t+1)y^2-2\right)\Bigr\} \nn \\
&\phantom{\times}
+\sum_{j=0}^{h-1}\sum_{r=2j}^{3j-1}\sum_{t=2(h-1-j)}^{3(h-1-j)-1}C_{j,r}C_{h-1-j,t}
(x^2-4)^{3h-2-r}(y^2-4)^{3h-2-t} \nn \\
&\phantom{\times\biggl[}\times(2r+1)(2t+1)\left((r+1)x^2+2\right)\left((t+1)y^2+2\right)\biggr]-\delta_{h0} 
\nn \\
\equiv&4(x^2-4)^{-3h-\frac12}(y^2-4)^{-3h-\frac12}P_h(x,y)-\delta_{h0},
\label{eq:explicitGammah}
\end{align}
where $P_h(x,y)$ is the symmetric polynomial of $x$ and $y$, 
because $\Gamma_h(x,y)=\Gamma_h(y,x)$. 

At first sight, it seems difficult to carry out the integrations with respect to $x$ and $y$ 
in \eqref{eq:two-ptintegral} 
because the denominator in \eqref{eq:two-ptandGamma} makes them coupled. 
However, it is shown in \cite{HT_2012} that there exists a two-point function 
at the same point 
\begin{align}
{}^{\exists}G_h(x,x)=\lim_{y\rightarrow x}G_h(x,y)
=\lim_{y\rightarrow x}\left(\frac{1}{2(x-y)^2}\Gamma_h(x,y)\right) 
\quad (h\in\bm Z_{\geq 0}). 
\label{eq:two-pointatone-point}
\end{align}
This implies that $\Gamma_h(x,y)$ can be divided by $(x-y)^2$. 
More precisely, for $h\in\bm N$, this equation means 
\begin{align}
{}^{\exists}G_h(x,x)
=&\lim_{y\rightarrow x}
\left(\frac{1}{2(x-y)^2}4(x^2-4)^{-3h-\frac12}(y^2-4)^{-3h-\frac12}P_h(x,y)\right) 
 \nn \\
=&2(x^2-4)^{-6h-1}\lim_{y\rightarrow x}\frac{P_h(x,y)}{(x-y)^2}. 
\label{eq:divisability}
\end{align}
Thus it follows that the polynomial $P_h(x,y)$ can be divided by $(x-y)^2$: 
$P_h(x,y)=(x-y)^2Q_h(x,y)$ 
with $Q_h(x,y)$ being a polynomial of $x$ and $y$ as well.\footnote
{In the case of $h=0$, it is true that $G_0(x,x)$ exists, 
and $P_0(x,y)$ defined in \eqref{eq:explicitGammah} is only the first order 
with respect to both $x$ and $y$. See eq.\eqref{eq:2pth=0}.} 
From this fact we find that the integrations of $x$ and $y$ are decoupled. 
This is the main observation in deriving the two-point function. 

\begin{figure}[h]
\centering
%\vspace{-1cm}
\includegraphics[width=10cm, bb=200 200 1000 600, clip]{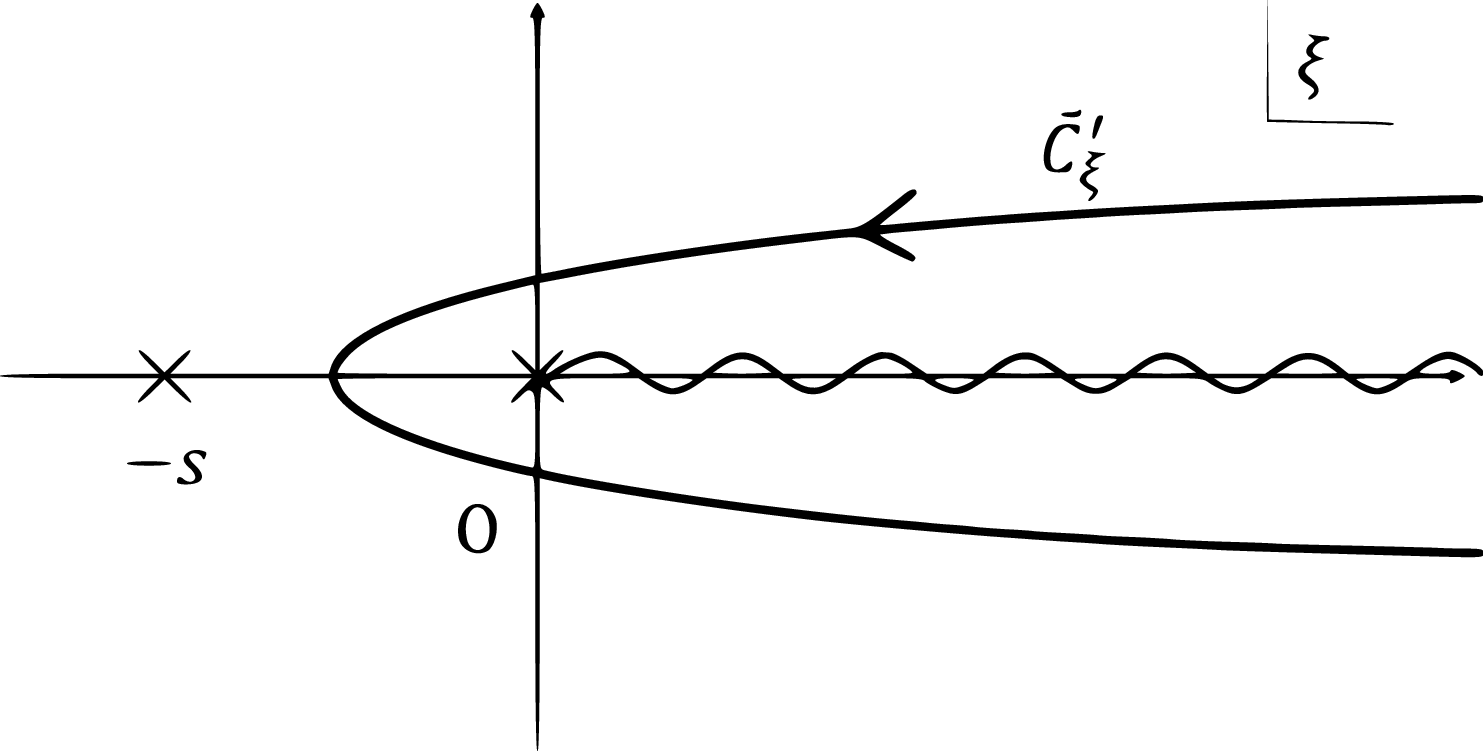}
%\vspace{-4cm}
\caption{\small Integration contour $\tilde{C}'_{\xi}$ on the complex $\xi$-plane.}
\label{fig:xicontour}
\end{figure}

Let us take a close look at how it works. Plugging \eqref{eq:two-ptandGamma} 
into \eqref{eq:two-ptintegral}, we get 
\begin{align}
&\left.\vev{\frac1N\tr f(\phi)\frac1N\tr g(\phi)}_{c,h}^{(1,0)}\right|_{\text{0-inst.}} \nn \\
&=\frac{1}{(2\pi i)^2}\oint_{\tilde C_x}dx\oint_{\tilde C_y}dy\,
f((\mu^2-x)^{\frac12})g((\mu^2-y)^{\frac12})\frac{1}{2(x-y)^2}\Gamma_h(x,y).
\label{eq:two-ptbyNicolai}
\end{align}  
For example, in $h=0$ case, by using \eqref{eq:Gamma0} this becomes 
\begin{align}
&\left.\vev{\frac1N\tr f(\phi)\frac1N\tr g(\phi)}_{c,0}^{(1,0)}\right|_{\text{0-inst.}} \nn \\
&=\frac{1}{(2\pi i)^2}\oint_{\tilde C_x}dx\oint_{\tilde C_y}dy\,
f((\mu^2-x)^{\frac12})g((\mu^2-y)^{\frac12})
\frac{1}{2(x-y)^2}\left(\frac{xy-4}{(x^2-4)^{\frac12}(y^2-4)^{\frac12}}-1\right).
\label{eq:2pth=0}
\end{align}
Then following the same derivation as in \cite{Kuroki:2012nt}, it is easy to see that 
for $f(\phi)=\phi^{2k+1}$ and $g(\phi)=\phi^{2\ell+1}$ this expression reproduces 
correctly the result there obtained by introducing the source term 
for the single trace operators.  For our purpose 
it is sufficient to consider \eqref{eq:two-ptbyNicolai} under the double scaling limit: 
\begin{align}
\mu^2=2+N^{-\frac23}s,\quad x=2+N^{-\frac23}\xi,\quad y=2+N^{-\frac23}\zeta.
\end{align}
We then find that for $h\in\bm Z_{\geq 0}$
\begin{align}
&\left.\vev{\frac1N\tr f(\phi)\frac1N\tr g(\phi)}_{c,h}^{(1,0)}\right|_{\text{0-inst.}} \nn \\
&=\frac{1}{(2\pi i)^2}\oint_{\tilde C'_{\xi}}d\xi\oint_{\tilde C'_{\zeta}}d\zeta\,
f\left(\left(N^{-\frac23}(s-\xi)\right)^{\frac12}\right)
g\left(\left(N^{-\frac23}(s-\zeta)\right)^{\frac12}\right)
\frac{2}{(\xi-\zeta)^2} \nn \\
&\times \Biggl[\xi^{-3h-\frac12}\zeta^{-3h-\frac12} 
\biggl\{\sum_{j=0}^h\sum_{r=2j}^{3j-1}\sum_{t=2(h-j)}^{3(h-j)-1}C_{j,r}C_{h-j,t}
(-4N^{-\frac23})^{-r-t-3}\xi^{3h-1-r}\zeta^{3h-1-t} \nn \\
&\phantom{\times \Biggl[4\xi^{-3h-\frac12}\zeta^{-3h-\frac12}\biggl\{}
\times\Bigl(2(2r+3)(2t+1)\xi+2(2r+1)(2t+3)\zeta\Bigr)N^{-\frac23} \nn \\
&\phantom{\times \Biggl[4\xi^{-3h-\frac12}\zeta^{-3h-\frac12}\biggl\{}
+4\sum_{j=0}^{h-1}\sum_{r=2j}^{3j-1}\sum_{t=2(h-1-j)}^{3(h-1-j)-1}
C_{j,r}C_{h-1-j,t}(-4N^{-\frac23})^{-r-t-5}\xi^{3h-2-r}\zeta^{3h-2-t} \nn \\
&\phantom{\times \Biggl[4\xi^{-3h-\frac12}\zeta^{-3h-\frac12}\biggl\{}
\times(2r+1)(2r+3)(2t+1)(2t+3)
\biggr\}-\delta_{h0}
\Biggr]\left(1+{\cal O}\left(N^{-\frac23}\right)\right),
\label{eq:two-ptfull}
\end{align} 
where $\tilde C'_{\xi}$ and $\tilde C'_{\zeta}$ are the contours 
in Fig.~\ref{fig:xicontour} on the complex $\xi$- and $\zeta$-plane, respectively. 
%%%%%%%%%%%%%%%%%%%%%%%%%%%%%%%%%%%%%%%%%%%%%%%%%%%%%%%%%%%%%%%%%%%%%%%
%%%%%%%%%%%%%%%%%%%%%%%%%%%%%%%%%%%%%%%%%%%%%%%%%%%%%%%%%%%%%%%%%%%%%%%%
We here note that in the sum over $r$ and $t$, 
a term with their largest value of $r$ and $t$ becomes the most dominant contribution 
under the double scaling limit due to the factor $(-4N^{-\frac23})^{-r-t-3}$ or 
$(-4N^{-\frac23})^{-r-t-5}$. Thus picking up the largest $r$ and $t$, we have  
\begin{align}
&\left.\vev{\frac1N\tr f(\phi)\frac1N\tr g(\phi)}_{c,h}^{(1,0)}\right|_{\text{0-inst.}} \nn \\
&=(-1)^{h+1}4^{-3h-1}\left(N^{-\frac23}\right)^{-3h}
\frac{1}{(2\pi i)^2}\oint_{\tilde C'_{\xi}}d\xi\oint_{\tilde C'_{\zeta}}d\zeta\,
f\left(\left(N^{-\frac23}(s-\xi)\right)^{\frac12}\right)
g\left(\left(N^{-\frac23}(s-\zeta)\right)^{\frac12}\right) \nn \\
&\phantom{=}\times\frac{2}{(\xi-\zeta)^2}\Biggl[\xi^{-3h-\frac12}\zeta^{-3h-\frac12}
\biggl\{\sum_{j=0}^hC_{j,3j-1}C_{h-j,3(h-j)-1}\xi^{3h-3j}\zeta^{3j} \nn \\
&\phantom{=\times\frac{2}{(\xi-\zeta)^2}\Biggl[\xi^{-3h-\frac12}\zeta^{-3h-\frac12}\biggl\{}
\times2\Bigl((6j+1)(6(h-j)-1)\xi+(6j-1)(6(h-j)+1)\zeta\Bigr) \nn \\
&\phantom{=\times\frac{2}{(\xi-\zeta)^2}\Biggl[\xi^{-3h-\frac12}\zeta^{-3}}
-16\sum_{j=0}^{h-1}C_{j,3j-1}C_{h-1-j,3(h-1-j)-1}\xi^{3(h-1-j)+2}\zeta^{3j+2} \nn \\
&\phantom{=\times\frac{2}{(\xi-\zeta)^2}
\Biggl[\xi^{-3h-\frac12}\zeta^{-3h-\frac12}\biggl\{}\times(36j^2-1)(36(h-1-j)^2-1)
\biggr\}-\delta_{h0}
\Biggr]\left(1+{\cal O}\left(N^{-\frac23}\right)\right). 
\label{eq:two-ptleading}
\end{align}
It is interesting that in the double scaling limit only $C_{j,3j-1}$ 
among $C_{j,r}$ ($2j\leq r\leq 3j-1$) contributes, which corresponds to 
the most singular term in the one-point function of the resolvent \eqref{eq:etah} 
at the edge of its cut $x=2$ where we take the double scaling limit. 
Hence this is also the case with the one-point fuctions of odd operators 
\cite{Kuroki:2016ucm}. 
This fact also makes it possible to find an explicit form of the two-point function, 
because these ``highest components'' $C_{j,3j-1}$ satisfy a closed recursion relation 
by themselves, which can be solved explicitly \cite{Kuroki:2016ucm} as 
\begin{align}
C_{j,3j-1}
=\frac{1}{4\sqrt{\pi}}\left(\frac{16}{3}\right)^j\frac{\Gamma\left(3j-\frac12\right)}{j!}.
\label{eq:highestcomp}
\end{align} 
Crucial observation in \eqref{eq:two-ptleading} is that the polynomial in the curly brackets 
is proportional to the leading order of $P_h(2+N^{-\frac23}\xi,2+N^{-\frac23}\zeta)$ 
in the $1/N$-expansion and hence for $h\in\bm N$, it must be divided by 
$(x-y)^2=N^{-\frac43}(\xi-\zeta)^2$, 
because $P_h(x,y)$ itself can be divided and is symmetric. 
Namely, there exits a symmetric polynomial $\gamma_h(\xi,\zeta)$ of $\xi$ and $\zeta$  
of order $3h-1$ such that 
\begin{align}
&2\sum_{j=0}^hC_{j,3j-1}C_{h-j,3(h-j)-1}\xi^{3h-3j}\zeta^{3j} 
\Bigl((6j+1)(6(h-j)-1)\xi+(6j-1)(6(h-j)+1)\zeta\Bigr) \nn \\
&-16\sum_{j=0}^{h-1}C_{j,3j-1}C_{h-1-j,3(h-1-j)-1}\xi^{3(h-1-j)+2}\zeta^{3j+2} 
(36j^2-1)(36(h-1-j)^2-1) \nn \\
&=(\xi-\zeta)^2\gamma_h(\xi,\zeta)
=(\xi-\zeta)^2
\sum_{\alpha,\beta\geq 0\atop \alpha+\beta=3h-1}
\gamma_{h,\alpha\beta}\xi^{\alpha}\zeta^{\beta} \qquad \qquad \qquad (h\in\bm N), 
\label{eq:smallgamma}
\end{align}
where $\gamma_{h,\alpha\beta}=\gamma_{h,\beta\alpha}$. 
For instance, 
\begin{align}
&\gamma_1(\xi,\zeta)=-5\zeta^2-3 \zeta\xi -5\xi^2, \nn \\
&\gamma_2(\xi,\zeta)
=-35(\zeta+\xi)\left(33\zeta^4-6\zeta^3\xi
+35\zeta^2\xi^2-6\zeta\xi^3+33\xi^4\right), \nn \\
&\gamma_3(\xi,\zeta)
=-70\left(12155\zeta^8+10725\zeta^7\xi+11011\zeta^6\xi^2+11066\zeta^5\xi^3
+10926\zeta^4\xi^4\right. \nn \\
&\phantom{\gamma_3(\xi,\zeta)=-70\left(12155\zeta^8\right.}
\left.+11066 \zeta^3\xi^5+11011\zeta^2\xi^6+10725\zeta\xi^7+12155\xi^8\right), 
\nn \\
&\cdots\, .
\label{eq:gammaex}
\end{align}
Substituting \eqref{eq:smallgamma} for \eqref{eq:two-ptleading}, we obtain 
\begin{align}
&\left.\vev{\frac1N\tr\phi^p\frac1N\tr\phi^q}_{c,h}^{(1,0)}\right|_{\text{0-inst.}} \nn \\
&=(-1)^{h+1}4^{-3h-1}\left(N^{-\frac23}\right)^{-3h}
\frac{1}{(2\pi i)^2}\oint_{\tilde C'_{\xi}}d\xi\oint_{\tilde C'_{\zeta}}d\zeta\,
f\left(\left(N^{-\frac23}(s-\xi)\right)^{\frac12}\right)
g\left(\left(N^{-\frac23}(s-\zeta)\right)^{\frac12}\right) \nn \\
&\phantom{=}\times\sum_{\alpha,\beta\geq 0\atop \alpha+\beta=3h-1}
2\gamma_{h,\alpha\beta}\xi^{-3h-\frac12+\alpha}\zeta^{-3h-\frac12+\beta}
\left(1+{\cal O}\left(N^{-\frac23}\right)\right) \qquad \qquad (h\in\bm N).
\label{eq:two-ptfactorizegeneral}
\end{align}

%%%%%%%%%%%%%%%%%%
\subsection{Odd-odd two-point function}
\label{subsec:odd-odd}
%%%%%%%%%%%%%%%%%%%%%%%%%%%%%%%%%%%%%%%%%%%%%%%%%%%%%%%%%%%
When $f(\phi)=\phi^p$, $g(\phi)=\phi^q$ for odd $p$ and $q$, 
eq.\eqref{eq:two-ptfactorizegeneral} becomes for $h\in\bm N$
\begin{align}
&\left.\vev{\frac1N\tr\phi^p\frac1N\tr\phi^q}_{c,h}^{(1,0)}\right|_{\text{0-inst.}} \nn \\
&=(-1)^{h+1}4^{-3h-1}\left(N^{-\frac23}\right)^{\frac{p+q}{2}-3h}
\sum_{\alpha,\beta\geq 0\atop \alpha+\beta=3h-1}2\gamma_{h,\alpha\beta} \nn \\
&\phantom{=}\times\frac{1}{2\pi i}\oint_{\tilde C'_{\xi}}d\xi\,
(s-\xi)^{\frac p2}\xi^{-3h-\frac12+\alpha}
\frac{1}{2\pi i}\oint_{\tilde C'_{\zeta}}d\zeta\,
(s-\zeta)^{\frac q2}\zeta^{-3h-\frac12+\beta}
\left(1+{\cal O}\left(N^{-\frac23}\right)\right).  
\label{eq:two-ptfactorize}
\end{align}
We thus have two decoupled integration each of which takes form of 
\begin{align}
I_{m,n}\equiv&\frac{1}{2\pi i}\oint_{\tilde C}dx\,(x^2-4)^{\frac m2}(\mu^2-x)^{\frac n2}
\nn \\
=&-\left(N^{-\frac23}\right)^{\frac{m+n}{2}+1}(-2i)^m
\frac{1}{2\pi i}\oint_{\tilde C'}d\xi\,\xi^{\frac{m}{2}}(s-\xi)^{\frac{n}{2}}
\left(1+{\cal O}(N^{-\frac23})\right), 
\label{eq:jmn}
\end{align}
where $\tilde C$ and $\tilde C'$ are the contours shown 
in Fig.~\ref{fig:xcontour} and Fig.~\ref{fig:xicontour}, respectively. 
As shown in \cite{Kuroki:2016ucm}, this is essentially the one-point function 
and when $m,n$ are odd, we have already found that
\begin{align}
I_{m,n}=
\left\{
\begin{array}{l}
\displaystyle{-\left(N^{-\frac23}\right)^{\frac{m+n}{2}+1}\frac{2^m}{\pi^2}
\frac{\Gamma\left(\frac m2+1\right)\Gamma\left(\frac n2+1\right)}
{\Gamma\left(\frac{m+n}{2}+2\right)}
s^{\frac{m+n}{2}+1}\ln s} \qquad (m+n\geq -2) \\
 \\
\displaystyle{\left(-N^{-\frac23}\right)^{\frac{m+n}{2}+1}
\frac{2^m}{\pi^2}
\Gamma\left(\frac m2+1\right)\Gamma\left(\frac n2+1\right)
\Gamma\left(-\frac{m+n}{2}-1\right)s^{\frac{m+n}{2}+1}} \nn \\
\phantom{\displaystyle{-\left(N^{-\frac23}\right)^{\frac{m+n}{2}+1}\frac{2^m}{\pi^2}
\frac{\Gamma\left(\frac m2+1\right)\Gamma\left(\frac n2+1\right)}
{\Gamma\left(\frac{m+n}{2}+2\right)}
s^{\frac{m+n}{2}+1}\ln s}\qquad} 
 (m+n<-2) 
\end{array}
\right..\\
\label{eq:Iresult}
\end{align}
Here we dropped less singular terms. Namely we have taken the most dominant 
nonanalytic term at $s=0$.\footnote
{As mentioned in \cite{Kuroki:2016ucm}, $I_{m,n}$ depends on $s$ 
only through combination $N^{-2/3}s$ due to $\mu^2=2+N^{-\frac23}s$. 
Therefore the most dominant term at $s=0$ have the largest power of $N$. } 
This means that we pick up a universal part of the correlation function 
which does not depend on detail of regularization \cite{Kuroki:2016ucm,Kuroki:2012nt}. 
Thus from \eqref{eq:two-ptfactorize} and \eqref{eq:jmn}, 
as far as the universal part is concerned, we finally arrive at a strikingly simple formula 
in the double scaling limit: for odd $p, q$ and $h\in\bm N$, 
\begin{align}
\left.\vev{\frac1N\tr\phi^p\frac1N\tr\phi^q}_{c,h}^{(1,0)}\right|_{\text{0-inst., univ.}}
=-8\sum_{\alpha,\beta\geq 0\atop \alpha+\beta=3h-1}\gamma_{h,\alpha\beta}\,
I_{-6h-1+2\alpha,p}\,I_{-6h-1+2\beta,q},
\label{eq:two-ptbyI}
\end{align}
As an example, \eqref{eq:gammaex}, \eqref{eq:Iresult}, and \eqref{eq:two-ptbyI} give 
the universal part of two-point functions of odd operators at genus one 
$\vev{\frac1N\tr\phi^{2k+1}\frac1N\tr\phi^{2\ell+1}}_{c,1}^{(1,0)}$ ($k\geq\ell$) as 
\begin{align}
&k=\ell=0: &&
\left.\vev{\frac1N\tr\phi\frac1N\tr\phi}_{c,1}^{(1,0)}\right|_{\text{0-inst., univ.}}
=-N^{\frac43}\frac{1}{48\pi^2}s^{-2}\ln s, \nn \\
&k=1,\,\ell=0: &&
\left.\vev{\frac1N\tr\phi^3\frac1N\tr\phi}_{c,1}^{(1,0)}\right|_{\text{0-inst., univ.}}
=N^{\frac23}\frac{1}{64\pi^2}s^{-1}\ln s, \nn \\
&k\geq 2,\,\ell=0: &&
\left.\vev{\frac1N\tr\phi^{2k+1}\frac1N\tr\phi}_{c,1}^{(1,0)}\right|_{\text{0-inst., univ.}}
=\left(N^{-\frac23}\right)^{k-2}\frac{1}{24\pi^3}
\frac{\Gamma\left(k+\frac32\right)\Gamma\left(\frac32\right)}{\Gamma(k-1)}
s^{k-2}\left(\ln s\right)^2, \nn \\
&k\geq 1,\,\ell=1: &&
\left.\vev{\frac1N\tr\phi^{2k+1}\frac1N\tr\phi^3}_{c,1}^{(1,0)}\right|_{\text{0-inst., univ.}}
=\left(N^{-\frac23}\right)^{k-1}\frac{k}{24\pi^3}
\frac{\Gamma\left(k+\frac32\right)\Gamma\left(\frac52\right)}{\Gamma(k)}
s^{k-1}\left(\ln s\right)^2,
 \nn \\
 &k\geq \ell\geq 2: &&
\left.\vev{\frac1N\tr\phi^{2k+1}\frac1N\tr\phi^{2\ell+1}}_{c,1}^{(1,0)} 
\right|_{\text{0-inst., univ.}} \nn \\
& && =\left(N^{-\frac23}\right)^{k+\ell-2}\frac{1}{24\pi^3}\left(k(k-1)+k\ell+\ell(\ell-1)\right) 
\frac{\Gamma\left(k+\frac32\right)\Gamma\left(\ell+\frac32\right)}
{\Gamma(k+1)\Gamma(\ell+1)}
s^{k+\ell-2}\left(\ln s\right)^2. \nn \\
\end{align}
It is worth pointing out that as in the case of the one-point function, we reconfirm that 
the double scaling limit works for the odd operators. 
Namely, \eqref{eq:Iresult} and \eqref{eq:two-ptbyI} implies that 
\begin{align}
\left.\vev{\frac1N\tr\phi^{2k+1}\frac1N\tr\phi^{2\ell+1}}_{c,h}^{(1,0)}
\right|_{\text{0-inst., univ.}}
\propto \left(N^{-\frac23}\right)^{-3h+1+k+\ell}.
\end{align}
Recalling the ``wave function renormalization" factor $\frac23(k+2)$ 
for the odd operator $\frac1N\tr\phi^{2k+1}$ mentioned below \eqref{eq:1ptgenusexp} 
and the normalization \eqref{eq:vev_MM}, in the whole two-point function 
$\left.\vev{\frac1N\tr\phi^{2k+1}\frac1N\tr\phi^{2\ell+1}}_c^{(1,0)}\right|_{\text{univ.}}$, 
the genus $h$ contribution takes the form 
\begin{align}
\left.\frac{1}{N^{2h+2}}
\vev{\frac1N\tr\phi^{2k+1}\frac1N\tr\phi^{2\ell+1}}_{c,h}^{(1,0)}
\right|_{\text{0-inst., univ.}}
N^{\frac23(k+2)}N^{\frac23(\ell+2)}\propto N^0
\label{eq:renormalization}
\end{align}
and hence each genus contribution becomes a function only of $s$ 
and contributes on an equal footing. Here we notice that 
the wave function renormalization does not change 
between the one-point and two-point functions 
because it is associated with the operator itself.  

In \cite{Kuroki:2012nt} we have already recognized that the two-point functions 
at genus zero for odd operators is expressed as a product of two hypergeometric functions 
each of which may have the logarithmic singular behavior. 
This is how the $(\ln s)^2$ appears in them. 
Now we find that this persists even at higher genus. 
In fact, \eqref{eq:two-ptbyI} implies that the two-point functions of odd operators 
at arbitrary genus are the sum of products of two $I_{m,n}$'s 
which are essentially the one-point functions with the possible 
$\ln s$ term as in \eqref{eq:Iresult}. 
Here \eqref{eq:two-ptbyI} should not be confused with the large-$N$ factorization 
because it refers to the connected part of the two-point correlation function.  
It would be interesting if \eqref{eq:two-ptbyI} can be derived independently 
by means of the Schwinger-Dyson equation of our SUSY double-well matrix model.

%%%%%%%%%%%%%%%%%%
\subsection{Higher genus contribution}
\label{subsec:large-h}
%%%%%%%%%%%%%%%%%%%%%%%%%%%%%%%%%%%%%%%%%%%%%%%%%%%%%%%%%%%
As we mentioned at the end of the previous section, eq.~\eqref{eq:two-ptbyI} tells us 
when the $\ln s$ term appears in the two-point function. 
We rewrite \eqref{eq:two-ptbyI} as 
\begin{align}
\left.\vev{\frac1N\tr\phi^p\frac1N\tr\phi^q}_{c,h}^{(1,0)}\right|_{\text{0-inst., univ.}}
=&-8\sum_{j=0}^{3h-1}\gamma_{h,j}\,
I_{-6h-1+2j,p}\,I_{-2j-3,q} \nn\\
&\text{with} \qquad \gamma_{h,j}\equiv\gamma_{h,j\,3h-1-j}=\gamma_{h,3h-1-j\,j},
\label{eq:two-ptbyI2}
\end{align}
and then from \eqref{eq:Iresult} the first factor $I_{-6h-1+2j,p}$ has $\ln s$ term 
when $j\geq 3h-\frac{p+1}{2}$, while the second one $I_{-2j-3,q}$ has 
when $j\leq\frac{q-1}{2}$. Hence the two-point function can involve 
$(\ln s)^2$ factor if and only if there exists $j$ such that 
$3h-\frac{p+1}{2}\leq j \leq\frac{q-1}{2}$. The necessary condition for the existence 
is $h\leq \frac{p+q}{6}$. Thus we arrive at an important conclusion that 
the $(\ln s)^2$ term appears only at lower genus depending on $p,q$. 
By using \eqref{eq:Iresult}, it is given by 
\begin{align}
\left.\vev{\frac1N\tr\phi^p\frac1N\tr\phi^q}_{c,h}^{(1,0)}\right|_{\text{0-inst., univ.}}
=&-\frac{1}{2\pi^4}\frac{1}{64^h}
\Gamma\left(\frac{p}{2}+1\right)\Gamma\left(\frac{q}{2}+1\right)
(N^{-\frac23}s)^{-3h+\frac{p+q}{2}}(\ln s)^2 \nn \\
&\times\sum_{j=3h-\frac{p+1}{2}}^{\frac{q-1}{2}}\gamma_{h,j}
\frac{\Gamma\left(-3h+j+\frac12\right)\Gamma\left(-j-\frac12\right)}
{\Gamma\left(-3h+j+\frac{p+3}{2}\right)\Gamma\left(-j+\frac{q+1}{2}\right)}.
\label{eq:lowergenus}
\end{align}
Note that the other terms in the sum on $j$ in \eqref{eq:two-ptbyI2} 
also have the same power of $N^{-\frac23}s$, 
but a lower power of $\ln s$, and therefore they are subleading. 

For the purpose of studying resurgence structure, 
we only need sufficiently higher genus contribution. 
Thus let us concentrate on the case $h>\frac{p+q}{6}$ where there is no $(\ln s)^2$ term. 
Since all terms in the sum over $j$ in \eqref{eq:two-ptbyI2} have the same 
power of $N^{-\frac23}s$, if there are terms with extra $\ln s$, 
they become the most dominant contribution at fixed $h$.  
Eq.~\eqref{eq:Iresult} implies that such terms appears when 
$3h-\frac{p+1}{2}\leq j\leq 3h-1$, or $0 \leq j\leq \frac{q-1}{2}$ and, therefore, 
\begin{align}
&\left.\vev{\frac1N\tr\phi^p\frac1N\tr\phi^q}_{c,h}^{(1,0)}
\right|_{\text{0-inst., univ.}} \nn \\
&=\frac{1}{2\pi^4}\frac{1}{64^h}
\Gamma\left(\frac{p}{2}+1\right)\Gamma\left(\frac{q}{2}+1\right)
(N^{-\frac23}s)^{-3h+\frac{p+q}{2}}\ln s \nn \\
&\phantom{=}\times\Biggl(
(-1)^{h+\frac{p+1}{2}}\sum_{j=0}^{\frac{q-1}{2}}(-1)^j\gamma_{h,j}
\frac{\Gamma\left(-j-\frac12\right)}{\Gamma\left(-j+\frac{q+1}{2}\right)}
\Gamma\left(-3h+j+\frac12\right)\Gamma\left(3h-j-\frac{p+1}{2}\right) \nn \\
&\phantom{==}+(-1)^{\frac{q-1}{2}}\sum_{j=3h-\frac{p+1}{2}}^{3h-1}(-1)^j\gamma_{h,j}
\Gamma\left(-j-\frac12\right)\Gamma\left(j-\frac{q-1}{2}\right)
\frac{\Gamma\left(-3h+j+\frac12\right)}{\Gamma\left(-3h+j+\frac{p+3}{2}\right)}
\Biggr) \nn \\
&=\frac{1}{2\pi^4}\left(-\frac{1}{64}\right)^h\Gamma\left(\frac{p}{2}+1\right)
\Gamma\left(\frac{q}{2}+1\right)
(N^{-\frac23}s)^{-3h+\frac{p+q}{2}}\ln s \nn \\
&\phantom{=}\times\Biggl(
(-1)^{\frac{p+1}{2}}\sum_{j=0}^{\frac{q-1}{2}}(-1)^j\gamma_{h,j}
\frac{\Gamma\left(-j-\frac12\right)}{\Gamma\left(-j+\frac{q+1}{2}\right)}
\Gamma\left(-3h+j+\frac12\right)\Gamma\left(3h-j-\frac{p+1}{2}\right) \nn \\
&\phantom{=\times\Biggl(}+(p\leftrightarrow q)\Biggr).
\label{eq:highergenus}
\end{align}
In appendix \ref{app:gamma} we give a formula of $\gamma_{h,j}$ and 
in principle we obtain the two-point function at each genus for any odd $p,q$ 
by plugging it into the above equation. However, in practice, $\gamma_{h,j}$ is too 
complicated to take the sum on $j$. In the next subsection, we argue that 
even if we cannot take the sum, still we can derive an explicit form of an ambiguity 
in the genus expansion from \eqref{eq:highergenus}.

%%%%%%%%%%%%%%%%%%%%%%%%%%%%%%%%%%%%%%%%%%
\subsection{Borel resummation}
Recalling \eqref{eq:vev_MM} and taking account of the wave function renormalization 
as in \eqref{eq:1ptgenusexp}, the genus expansion 
of the two-point function of the odd operators in the zero-instanton sector is given by  
\begin{align}
&N^{\frac{p+q}{3}+2}
\left.\vev{\frac1N\tr\phi^p\frac1N\tr\phi^q}_c^{(1,0)}\right|_{\text{0-inst., univ.}}
=N^{\frac{p+q}{3}+2}\sum_{h=0}^{\infty}\frac{1}{N^{2h+2}}
\left.\vev{\frac1N\tr\phi^p\frac1N\tr\phi^q}_{c,h}^{(1,0)}\right|_{\text{0-inst., univ.}}. 
\end{align}
Here the sum on $h$ is classified into lower genus contribution with $h\leq\frac{p+q}{6}$ 
and higher genus one with $h>\frac{p+q}{6}$. In the former, we have the $(\ln s)^2$ term 
as in \eqref{eq:lowergenus} and in the latter, only the $\ln s$ term appears 
as in \eqref{eq:highergenus}. The former is only a finite sum without any ambiguity, 
while the latter is an infinite sum and, as we will see later, 
it is non-Borel summable with ambiguity. 
Thus hereafter we concentrate on the higher genus contribution 
which reads from \eqref{eq:highergenus} as 
\begin{align}
&N^{\frac{p+q}{3}+2}
\left.\vev{\frac1N\tr\phi^p\frac1N\tr\phi^q}_c^{(1,0)}\right|_{\text{0-inst., univ.}} \nn \\
&=\frac{1}{2\pi^4}\Gamma\left(\frac p2+1\right)\Gamma\left(\frac q2+1\right)
s^{\frac{p+q}{2}}\ln s\sum_{h=\mathrm{ceil}\left(\frac{p+q}{6}\right)}^{\infty}
\left(-\frac{1}{64s^3}\right)^h(-1)^{\frac{p+1}{2}} \nn \\
&\phantom{=}\times\sum_{j=0}^{\frac{q-1}{2}}(-1)^j\gamma_{h,j}
\frac{\Gamma\left(-j-\frac12\right)}{\Gamma\left(-j+\frac{q+1}{2}\right)}
\Gamma\left(-3h+j+\frac12\right)\Gamma\left(3h-j-\frac{p+1}{2}\right) \nn \\
&\phantom{=}+(p\leftrightarrow q)+(\text{finite sum}),
\label{eq:genusexp}
\end{align}
where $\mathrm{ceil}(a)$ is the least integer that is greater than or equal to $a$. 
For the same reason, as far as ambiguity is concerned, we have only to take care 
of the sum on $h$ from $h\gg 1$ to $\infty$:  @
\begin{align}
&N^{\frac{p+q}{3}+2}
\left.\vev{\frac1N\tr\phi^p\frac1N\tr\phi^q}_c^{(1,0)}\right|_{\text{0-inst., univ.}} \nn \\
&=\frac{1}{2\pi^4}\Gamma\left(\frac p2+1\right)\Gamma\left(\frac q2+1\right)
s^{\frac{p+q}{2}}\ln s\sum_{h\gg 1}^{\infty}
\left(-\frac{1}{64s^3}\right)^h(-1)^{\frac{p+1}{2}} \nn \\
&\phantom{=}\times\sum_{j=0}^{\frac{q-1}{2}}(-1)^j\gamma_{h,j}
\frac{\Gamma\left(-j-\frac12\right)}{\Gamma\left(-j+\frac{q+1}{2}\right)}
\Gamma\left(-3h+j+\frac12\right)\Gamma\left(3h-j-\frac{p+1}{2}\right) \nn \\
&\phantom{=}+(p\leftrightarrow q)+(\text{finite sum}),
\label{eq:replacedgenusexp}
\end{align}
where the sum on not large $h$ is included in $(\text{finite sum})$ term. 
In order to find large order behavior of this genus expansion, 
we need behavior of $\gamma_{h,j}$. Here it is sufficient to use the fact that 
\begin{align}
\gamma_{h,j}=\frac{\Gamma\left(3h+\frac12-j\right)}{h!}\left(\frac{16}{3}\right)^hf_j(h),
\label{eq:fjdef}
\end{align}
where $f_j(h)$ is a polynomial of $h$ of degree $j$ as 
\begin{align}
f_j(h)=-\frac{3^j}{2\sqrt{\pi}}h^j+\frac{3^{j-1}}{4\sqrt{\pi}}(j^2+2)h^{j-1}+{\cO}(h^{j-2}).
\label{eq:fjform}
\end{align}
These properties of $\gamma_{h,j}$ are proved in appendix \ref{app:gammaprop}. 
Plugging \eqref{eq:fjdef} into \eqref{eq:replacedgenusexp}, we get 
\begin{align}
&N^{\frac{p+q}{3}+2}
\left.\vev{\frac1N\tr\phi^p\frac1N\tr\phi^q}_c^{(1,0)}\right|_{\text{0-inst., univ.}} \nn \\
&=\frac{1}{2\pi^4}\Gamma\left(\frac p2+1\right)\Gamma\left(\frac q2+1\right)
s^{\frac{p+q}{2}}\ln s\sum_{h\gg 1}^{\infty}
\left(-\frac{1}{64s^3}\right)^hS_{pq}(h)+(p\leftrightarrow q)+(\text{finite sum}), 
 \label{eq:Sdef} 
\end{align}
where
\begin{align}
&S_{pq}(h)\equiv(-1)^{\frac{p+1}{2}}
\sum_{j=0}^{\frac{q-1}{2}}(-1)^j\gamma_{h,j}
\frac{\Gamma\left(-j-\frac12\right)}{\Gamma\left(-j+\frac{q+1}{2}\right)}
\Gamma\left(-3h+j+\frac12\right)\Gamma\left(3h-j-\frac{p+1}{2}\right) \nn \\
&=(-1)^{\frac{p+1}{2}}
\sum_{j=0}^{\frac{q-1}{2}}(-1)^j\frac{\Gamma\left(3h+\frac12-j\right)}{h!}
\left(\frac{16}{3}\right)^hf_j(h) \nn \\
&\phantom{(-1)^{\frac{p+1}{2}}\sum(-1)^j}
\times \frac{\Gamma\left(-j-\frac12\right)}{\Gamma\left(-j+\frac{q+1}{2}\right)}
\Gamma\left(-3h+j+\frac12\right)\Gamma\left(3h-j-\frac{p+1}{2}\right) \nn \\
&=(-1)^{\frac{p+1}{2}}\pi\left(-\frac{16}{3}\right)^h\sum_{j=0}^{\frac{q-1}{2}}
\frac{\Gamma\left(-j-\frac12\right)}{\Gamma\left(-j+\frac{q+1}{2}\right)}
\frac{\Gamma\left(3h-j-\frac{p+1}{2}\right)}{h!}f_j(h).
\label{eq:Spq}
\end{align}
Combining \eqref{eq:Sdef} and \eqref{eq:Spq}, 
we find that \eqref{eq:Sdef} is a positive term series 
whose large order behavior is stringy: $\frac{\Gamma(3h-j-\frac{p+1}{2})}{h!}\sim (2h)!$. 
This is very similar to the one-point function \eqref{eq:1ptgenusexp} and 
provides further support that our matrix model describes a string theory 
in the double scaling limit \cite{Shenker:1990uf}. 
In fact, the one-point function is given as \cite{Kuroki:2016ucm} 
\begin{align}
\vev{\frac1N\tr\phi^p}_h^{(1,0)}=C_{h,3h-1}I_{-6h+1,p}
\end{align}
for odd $p$. The expression of $I_{m,n}$ in \eqref{eq:Iresult} implies that 
it does not have factorial growth associated with $m$ 
because of $\frac{\Gamma\left(\frac m2+1\right)}
{\Gamma\left(\frac{m+n}{2}+2\right)}$ ($m+n\geq -2$) or 
$\Gamma\left(\frac m2+1\right)\Gamma\left(-\frac{m+n}{2}-1\right)$ ($m+n<-2$). 
Thus it is $C_{h,3h-1}$ that provides $(2h)!$ growth as in \eqref{eq:highestcomp}. 
This is also the case with the two-point function given in \eqref{eq:two-ptbyI2}
where two $I$'s do not grow as $(2h)!$, while $\gamma_{h,j}$ does due to \eqref{eq:fjdef}. 

\begin{figure}[h]
\centering
\includegraphics[width=10cm, bb=0 0 800 800, clip]{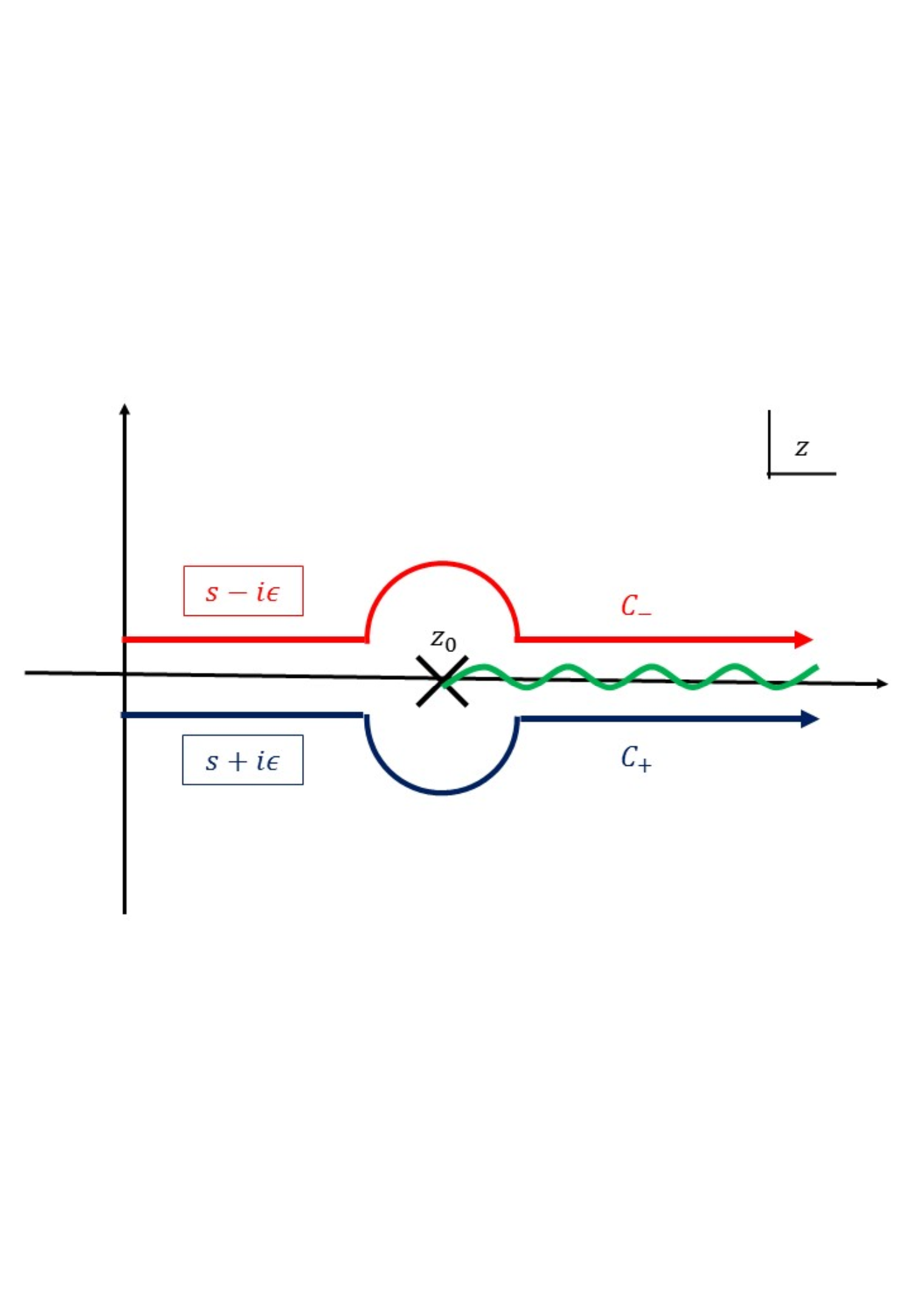}
\vspace{-4cm}
\caption{\small Integration contours $C_+$ and $C_-$ on the Borel plane.}
\label{fig:twocontours}
\end{figure}

Thus \eqref{eq:Sdef} is a divergent series with convergence radius zero.  
In order to make it well-defined, let us apply the Borel resummation technique 
to \eqref{eq:Sdef}. It amounts to inserting  
\begin{align}
1=\frac{1}{\Gamma\left(2h+1\right)}\int_0^\infty dz\,z^{2h}e^{-z}
\label{eq:resummationidentity}
\end{align}
into \eqref{eq:Sdef} and interchanging the order of the sum on $h$ 
and the integration on $z$. Then 
\begin{align}
&N^{\frac{p+q}{3}+2}
\left.\vev{\frac1N\tr\phi^p\frac1N\tr\phi^q}_c^{(1,0)}\right|_{\text{0-inst., univ., resum}} 
 \nn \\
&=\frac{1}{2\pi^4}
\Gamma\left(\frac p2+1\right)\Gamma\left(\frac q2+1\right)
s^{\frac{p+q}{2}}\ln s\int_{0}^{\infty}dz\,e^{-z}T_{pq}(z)
+(p\leftrightarrow q)+(\text{finite sum}), 
 \label{eq:Tdef} \\
&T_{pq}(z)\equiv\sum_{h\gg 1}^{\infty}\frac{1}{\Gamma(2h+1)}S_{pq}(h)
\left(-\frac{z^2}{64s^3}\right)^h. 
\end{align}
Using \eqref{eq:Spq}, $T_{pq}(z)$ becomes 
\begin{align}
T_{pq}(z)=&(-1)^{\frac{p+1}{2}}\pi\sum_{j=0}^{\frac{q-1}{2}}
\frac{\Gamma\left(-j-\frac12\right)}{\Gamma\left(-j+\frac{q+1}{2}\right)}
\sum_{h\gg 1}
\frac{\Gamma\left(3h-j-\frac{p+1}{2}\right)}{\Gamma(2h+1)h!}
f_j(h)\left(\frac{z^2}{12s^3}\right)^h.
\label{eq:T}
\end{align}
Applying the Stirling formula, we have for $\ell\in\bm N$ and large $h$
\begin{align}
&\frac{\Gamma(3h-\ell)}{h!\Gamma(2h+1)} \sim 
\frac{1}{2\sqrt{\pi}\,3^{\ell+\frac{1}{2}}}\left(\frac{27}{4}\right)^h h^{-\ell-\frac32}
\left[1+\left\{\ell(\ell+1)-\frac{7}{12}\right\}\frac{1}{6h} +\cO(h^{-2})\right] \nn \\
&=\frac{(-1)^{h+\ell+1}\sqrt{\pi}}{2\cdot 3^{\ell+\frac{1}{2}}}\left(\frac{27}{4}\right)^h
\biggl[\frac{1}{\Gamma(\ell+\frac32)} \binomi{\ell+\frac12}{h} \nn \\
&\phantom{=\frac{(-1)^{h+\ell+1}\sqrt{\pi}}{2\cdot 3^{\ell+\frac{1}{2}}}
\left(\frac{27}{4}\right)^h\biggl[}
+\frac{12\ell^2+30\ell+17}{36}\,\frac{1}{\Gamma(\ell+\frac52)} \binomi{\ell+\frac32}{h} 
+\cO(h^{-\ell-\frac{7}{2}})\biggr].
\label{eq:Striling}
\end{align}
We here note that since $\binom{\alpha+\frac12}{h}$ is of $\cO(h^{-\alpha-\frac32})$ 
for $\alpha\in\bm N$, this is just change of a base in terms of which 
a function of $h$ is expanded. Thus $T_{pq}(z)$ can be rewritten as 
\begin{align}
T_{pq}(z)\sim&-\frac{\pi^{\frac32}}{2\cdot3^{\frac p2+1}}
\sum_{j=0}^{\frac{q-1}{2}}\left(-\frac13\right)^j
\frac{\Gamma\left(-j-\frac12\right)}{\Gamma\left(-j+\frac{q+1}{2}\right)}
\frac{1}{\Gamma\left(j+\frac p2+2\right)} \nn \\
&\times\sum_{h\gg 1}^{\infty}
\Biggl[
\binom{j+\frac p2+1}{h}+\cO(h^{-j-\frac p2-3})
\Biggr]
f_j(h)\left(-\frac{9z^2}{16s^3}\right)^h.
\label{eq:Tbybinom}
\end{align}
In order to take the sum over $h$, we utilize an identity 
\begin{align}
f_j(h)\binomi{\alpha}{h}=\sum_{k=0}^jb_k^{(j)}(\alpha)\binom{\alpha-k}{h-k} 
\end{align}
with $b_k^{(j)}(\alpha)$ independent of $h$ which is shown in appendix \ref{app:gammaprop}. 
Then we obtain 
\begin{align}
&T_{pq}(z)\sim -\frac{\pi^{\frac32}}{2\cdot3^{\frac p2+1}}
\sum_{j=0}^{\frac{q-1}{2}}\left(-\frac13\right)^j
\frac{\Gamma\left(-j-\frac12\right)}{\Gamma\left(-j+\frac{q+1}{2}\right)}
\frac{1}{\Gamma\left(j+\frac p2+2\right)} \nn \\
&\times\sum_{h\gg 1}^{\infty}
\Biggl[
\sum_{k=0}^jb_k^{(j)}\left(j+\frac p2+1\right)
\binom{j+\frac p2+1-k}{h-k}+\cO(h^{-j-\frac p2-3})
\Biggr]
\left(-\frac{9z^2}{16s^3}\right)^h,
\end{align}
and by noting an identity
\begin{align}
\sum_{h=0}^{\infty}\binom{\alpha-k}{h-k}x^h=x^k(1+x)^{\alpha-k}
=\sum_{\ell=0}^k(-1)^{\ell}\binom{k}{\ell}(1+x)^{\alpha-\ell},
\end{align}
we can take the sum on $h$ as 
\begin{align}
T_{pq}(z)\sim&-\frac{\pi^{\frac32}}{2\cdot3^{\frac p2+1}}
\sum_{j=0}^{\frac{q-1}{2}}\left(-\frac13\right)^j
\frac{\Gamma\left(-j-\frac12\right)}{\Gamma\left(-j+\frac{q+1}{2}\right)}
\frac{1}{\Gamma\left(j+\frac p2+2\right)} \nn \\
&\times
\sum_{k=0}^jb_k^{(j)}\left(j+\frac p2+1\right)
\sum_{\ell=0}^k(-1)^{\ell}\binom{k}{\ell}\left(1-\frac{z^2}{z_0^2}\right)^{j+\frac p2+1-\ell} 
+\cO\left(\left(1-\frac{z^2}{z_0^2}\right)^{\frac p2+2}\right) \nn \\
&+(\text{finite sum}),
\label{eq:Tafterhsum}
\end{align}
where we extend the sum of $h$ from $0$ to $\infty$ because difference is again 
only a finite sum without any ambiguity, and set
\begin{align}
z_0\equiv\frac43s^{\frac32}.
\label{eq:z0def}
\end{align}
As shown in \eqref{eq:Tdef}, in order to obtain the Borel resummation 
of the two-point function, it is necessary to evaluate the integral 
\begin{align}
\int_0^{\infty}dz\,e^{-z}T_{pq}(z), 
\end{align}
and from \eqref{eq:Tafterhsum}, it amounts to considering 
\begin{align}
u_{\alpha}(s)\equiv\int_0^{\infty}dz\,e^{-z}\left(1-\frac{z^2}{z_0^2}\right)^{\alpha}.   
\label{eq:Bessel}
\end{align}
Now we recognize that the two-point function \eqref{eq:Tdef} is non-Borel summable, 
because the integrand in \eqref{eq:Bessel} has a cut from $z=z_0$ along the positive 
real axis for $\alpha\notin\bm Z$ and we have two ways to bypass it as shown 
in Fig.~\ref{fig:twocontours}. 
%%%%%%%%%%%%%%%%%%%%%%%%%%%%%%%%%%%%%%%%%%%%%%%%%%%%%%%%%%%%%%%%%%%%%%%% 
%%%%%%%%%%%%%%%%%%%%%%%%%%%%%%%%%%%%%%%%%%%%%%%%%%%%%%%%%%%%%%%%%%%%%%%%
We identify an ambiguity as difference between them:  
\begin{align}
\text{Amb.}\,u_{\alpha}(s)\equiv& u_{\alpha}(s+i\epsilon)-u_{\alpha}(s-i\epsilon)
=\int_{C_+}dz\,e^{-z}\left(1-\frac{z^2}{z_0^2}\right)^{\alpha}
-\int_{C_-}dz\,e^{-z}\left(1-\frac{z^2}{z_0^2}\right)^{\alpha} \nn \\
=&2i\sin(\alpha\pi)\int_{z_0}^{\infty}dz\,e^{-z}\left(\frac{z^2}{z_0^2}-1\right)^{\alpha} 
 \nn \\
=&\frac{i}{\sqrt{\pi}}\sin(\alpha\pi)\frac{3^{\alpha-\frac12}}{2^{\alpha-\frac52}}
s^{-\frac32\left(\alpha-\frac12\right)}\Gamma(\alpha+1)K_{\alpha+\frac12}(z_0) 
 \nn \\
\sim&2i\sin(\alpha\pi)\Gamma(\alpha+1)\left(\frac32\right)^{\alpha}
s^{-\frac32\alpha}e^{-\frac43s^{\frac32}}
\left(1+\frac38\alpha(\alpha+1)s^{-\frac32}+\cO(s^{-3})\right),
\end{align}
where we have used the asymptotic form of the modified Bessel function. 
Important observation here is that $\text{Amb.}\,u_{\alpha}(s)$ 
with the smallest $\alpha$ becomes 
the most dominant in the large-$s$ regime. Hence picking up the smallest power 
of $\left(1-\frac{z^2}{z_0^2}\right)$ in \eqref{eq:Tafterhsum}, we find 
\begin{align}
\text{Amb.}\int_0^{\infty}dz\,e^{-z}T_{pq}(z)
\sim&-\frac{\pi^{\frac32}}{2\cdot3^{\frac p2+1}}
\sum_{j=0}^{\frac{q-1}{2}}\left(-\frac13\right)^j
\frac{\Gamma\left(-j-\frac12\right)}{\Gamma\left(-j+\frac{q+1}{2}\right)}
\frac{1}{\Gamma\left(j+\frac p2+2\right)} \nn \\
&\times
b_j^{(j)}\left(j+\frac p2+1\right)(-1)^j\,\text{Amb.}\,u_{\frac p2+1}(s) 
\left(1+\cO\left(s^{-\frac32}\right)\right). 
\end{align}
Finally we use the result shown in appendix \ref{app:gammaprop}
\begin{align}
&b_j^{(j)}(\alpha)=-\frac{3^j}{2\sqrt{\pi}}\frac{\Gamma(\alpha+1)}{\Gamma(\alpha+1-j)}, 
\end{align}
and then from \eqref{eq:Tdef}, we obtain 
\begin{align}
&\text{Amb.}\,N^{\frac{p+q}{3}+2}
\left.\vev{\frac1N\tr\phi^p\frac1N\tr\phi^q}_c^{(1,0)}
\right|_{\text{0-inst., univ.}} \nn \\
&=i\sin\left(\frac{p}{2}\pi\right)\frac{1}{2^{\frac p2+2}\pi^{\frac52}q}
\Gamma\left(\frac p2+1\right)
\frac{\Gamma\left(\frac q2+1\right)}{\Gamma\left(\frac{q+1}{2}\right)}
s^{-\frac{p}{4}+\frac{q}{2}-\frac32}e^{-\frac43s^{\frac32}}\ln s
\left(1+\cO(s^{-\frac32})\right) +(p\leftrightarrow q),
\label{eq:ambforallpq}
\end{align}
where the sum on $j$ is taken as   
\begin{align}
\sum_{j=0}^{\frac{q-1}{2}}
\frac{\Gamma\left(-j-\frac12\right)}{\Gamma\left(-j+\frac{q+1}{2}\right)}
=-\frac{2\sqrt{\pi}}{q\Gamma\left(\frac{q+1}{2}\right)}.
\end{align}
The above derivation implies that all $j$'s in the sum in \eqref{eq:two-ptbyI2} 
contribute to the leading order in the ambiguity. 
Without loss of generality we can assume $p\leq q$ and then 
\eqref{eq:ambforallpq} leads to 
\begin{align}
&\text{Amb.}\,N^{\frac{p+q}{3}+2}
\left.\vev{\frac1N\tr\phi^p\frac1N\tr\phi^q}_c^{(1,0)}
\right|_{\text{0-inst., univ.}} \nn \\
&=i(1+\delta_{pq})\sin\left(\frac{p}{2}\pi\right)\frac{1}{2^{\frac p2+2}\pi^{\frac52}q}
\Gamma\left(\frac p2+1\right)
\frac{\Gamma\left(\frac q2+1\right)}{\Gamma\left(\frac{q+1}{2}\right)}
s^{-\frac{p}{4}+\frac{q}{2}-\frac32}e^{-\frac43s^{\frac32}}\ln s
\left(1+\cO(s^{-\frac32})\right). 
\label{eq:0-instamb}
\end{align}
As in \eqref{eq:Z_pinst_dsl} it is shown that weight of the one-instanton 
in our matrix model is proportional to $e^{-\frac43s^{\frac32}}$ 
and, therefore, it is likely that 
the ambiguity in the zero-instanton sector in \eqref{eq:0-instamb} would be 
canceled by another ambiguity in the one-instanton sector according to 
resurgence. We will confirm this in the next section.

%%%%%%%%%%%%%%%%%%%%%%%%%%%%%%%%%%%%%%%%%%%%%%%%%%%%%%%%%%%%%%%%%%%%%%%%%%%%%
\section{Two-point function in the one-instanton sector}
\label{sec:one-inst}
\setcounter{equation}{0}
%%%%%%%%%%%%%%%%%%%%%%%%%%%%%%%%%%%%%%%%%%%%%%%%%%%%%%%%%%%%%%%%%%%%%%%%%%%%%
In the previous section, we derived the perturbative series of the two-point function 
of the odd operators under the double scaling limit. We found that it is non-Borel 
summable and gave the explicit form of its ambiguity at the leading order of the large-$s$ 
regime. In this section we consider the two-point function in the one-instanton sector 
and show that it also has ambiguity, which exactly cancels that in the zero-instanton 
sector at the leading order in the large-$s$ regime. 
Thus we confirm that resurgence works in the two-point functions. 

%%%%%%%%%%%%%%%%%%%%%%%%%%%%%%%%%%%%%%%%
\subsection{Review of two-point function in the random matrix theory}
As in \eqref{eq:two-ptintegral}, the two-point function in our model can be deduced 
from that in the Gaussian matrix model, where a nice formula for the two-point function 
has been already known in the literature. Hence we review it in the context 
of the Gaussian Unitary Ensemble (GUE). 

The GUE is defined in terms of the partition function
\begin{align}
Z^{(G)}\equiv\int dM\,e^{-\frac{N}{2}\tr M^2}
=\tilde C_N\int_{-\infty}^{\infty}\prod_{i=1}^Ndx_i\,
\Delta(x)^2\,e^{-N\sum_i\frac12x_i^2},
\label{eq:GUE}
\end{align} 
where $M$ is an $N\times N$ Hermitian matrix and $x_i$ ($i=1,\cdots, M$) 
are its eigenvalues. Note that in this section we discuss the standard GUE 
in \eqref{eq:GUE}, which is different from the one in \eqref{eq:modifiedGaussian} 
with the upper bound for $x_i$ we have already discussed. 
Associated with \eqref{eq:GUE}, a joint probability distribution is defined as 
\begin{align}
P(\{x\})\equiv\frac{\tilde C_N}{Z^{(G)}}\Delta(x)^2\,e^{-N\sum_i\frac12x_i^2},
\end{align}
then 
\begin{align}
\int\prod_{i=1}^Ndx_i\,P({\{x\}})=1. 
\end{align}
The two-point function, or covariance in the GUE is given as 
\begin{align}
&\vev{\frac1N\tr f(M)\frac1N\tr g(M)}^{(G)}
=\frac{1}{Z^{(G)}}\int dM\,\frac1N\tr f(M)\frac1N\tr g(M)\,e^{-\frac{N}{2}\tr M^2} \nn \\
&=\frac{\tilde C_N}{Z^{(G)}}\int_{-\infty}^{\infty}\prod_{i=1}^Ndx_i\,\Delta(x)^2
\frac{1}{N^2}\sum_{i,j=1}^Nf(x_i)g(x_j)\,e^{-N\sum_i\frac12x_i^2} \nn \\
&=\frac{\tilde C_N}{Z^{(G)}}\int_{-\infty}^{\infty}\prod_{i=1}^Ndx_i\,\Delta(x)^2
\frac{1}{N^2}(Nf(x_1)g(x_1)+N(N-1)f(x_1)g(x_2))
\,e^{-N\sum_i\frac12x_i^2}. 
\end{align}
Introducing the $k$-point correlation function obtained by integrating 
$P(\{x\})$ with respect to $N-k$ variables 
\begin{align}
R_k(x_1,\cdots,x_k)\equiv\frac{N!}{(N-k)!}\int_{-\infty}^{\infty}\prod_{i=k+1}^Ndx_i\,
P(\{x\}),
\end{align}
the two-point function becomes 
\begin{align}
&\vev{\frac1N\tr f(M)\frac1N\tr g(M)}^{(G)} \nn \\
&=\frac{1}{N^2}\int_{-\infty}^{\infty} dx\,f(x)g(x)R_1(x)
+\frac{1}{N^2}\int_{-\infty}^{\infty} dxdy\,f(x)g(y)R_2(x,y).
\label{eq:two-ptbyR}
\end{align}
Now an important role is played by a kernel 
\begin{align}
K(x,y)=&e^{-\frac{N}{4}\left(x^2+y^2\right)}\sum_{n=0}^{N-1}\frac{1}{h_n}p_n(x)p_n(y) \nn \\
=&e^{-\frac{N}{4}\left(x^2+y^2\right)}\frac{1}{h_{N-1}}
\frac{p_N(x)p_{N-1}(y)-p_{N-1}(x)p_N(y)}{x-y},
\label{eq:K}
\end{align}
where $p_n(x)$ ($n=0,\cdots,N-1$) is a monic orthogonal polynomial of degree $n$ 
satisfying  
\begin{align}
\int_{-\infty}^{\infty}dx\,e^{-\frac{N}{2}x^2}p_m(x)p_n(y)=h_n\delta_{mn}. 
\end{align}
More precisely, 
\begin{align}
&p_n(x)=\frac{1}{(2N)^{\frac{n}{2}}}H_n\left(\sqrt{\frac{N}{2}}x\right), \qquad 
\text{with} \quad 
H_n(x)=(-1)^ne^{x^2}\frac{d^n}{dx^n}e^{-x^2}, \nn \\
&h_n=\frac{\sqrt{2\pi}n!}{N^{n+\frac12}}. 
\end{align}
The kernel is related to the one-point and two-point functions as 
\begin{align}
&R_1(x)=K(x,x)=N\rho(x), \qquad \text{with} \quad 
\rho(x)=\vev{\frac1N\delta(x-M)}^{(G)}, \nn \\
&R_2(x,y)=K(x,x)K(y,y)-K(x,y)^2.
\label{eq:RandK}
\end{align}
Substituting these equations for \eqref{eq:two-ptbyR} yields 
\begin{align}
&\vev{\frac1N\tr f(M)\frac1N\tr g(M)}^{(G)}=\vev{\frac1N\tr f(M)g(M)}^{(G)} \nn \\
&+\vev{\frac1N\tr f(M)}^{(G)}\vev{\frac1N\tr g(M)}^{(G)}
-\frac{1}{N^2}\int_{-\infty}^{\infty} dxdy\,f(x)g(y)K(x,y)^2.
\end{align}
Therefore, 
\begin{align}
\vev{\frac1N\tr f(M)\frac1N\tr g(M)}^{(G)}_c
=\vev{\frac1N\tr f(M)g(M)}^{(G)}-\frac{1}{N^2}\int_{-\infty}^{\infty} dxdy\,f(x)g(y)K(x,y)^2.
\label{eq:GUEtwo-pt}
\end{align}

%%%%%%%%%%%%%%%%%%%%%%%%%%%%%%%%%%%%%%%%
\subsection{Application to our model}
In our model, as in \eqref{eq:resolventbyNicolai}, 
the two-point function of odd operators in the filling fraction $(1,0)$ is expressed 
via the Nicolai mapping as  
\begin{align}
\vev{\frac1N\tr\phi^p\frac1N\tr\phi^q}^{(1,0)}
=&\frac{\tilde C_N}{Z_{(1,0)}}\int_0^{\infty}
\prod_{k=1}^N\left(2\lambda_kd\lambda_k\right)\Delta(\lambda^2)^2
\frac{1}{N^2}\sum_{i,j=1}^N\lambda_i^p\lambda_j^q\,
e^{-\frac{N}{2}\sum_k(\lambda_k^2-\mu^2)^2} \nn \\
=&\frac{\tilde C_N}{Z_{(1,0)}}\int_{-\infty}^{\mu^2}
\prod_{k=1}^Ndx_k\,\Delta(x)^2
\frac{1}{N^2}\sum_{i,j=1}^N(\mu^2-x_i)^{\frac{p}{2}}(\mu^2-x_j)^{\frac{q}{2}}\,
e^{-\frac{N}{2}\sum_kx_k^2} \nn \\
=&\vev{\frac1N\tr(\mu^2-M)^{\frac{p}{2}}\frac1N\tr(\mu^2-M)^{\frac{q}{2}}}^{(G')},
\label{eq:twp-ptNicolai}
\end{align}
where $(G')$ indicates that $x_i$ integration has upper boundary $x_i=\mu^2$ 
as in \eqref{eq:modifiedGaussian}. Since \eqref{eq:RandK} follows from existence 
of the orthogonal polynomials, in the GUE with the boundary we also have a formula 
of the two-point function similar to \eqref{eq:GUEtwo-pt}: 
\begin{align}
\vev{\frac1N\tr f(M)\frac1N\tr g(M)}^{(G')}_c
=\vev{\frac1N\tr f(M)g(M)}^{(G')}-\frac{1}{N^2}\int_{-\infty}^{\mu^2} dxdy\,f(x)g(y)
K^{(G')}(x,y)^2,
\label{eq:G'two-pt}
\end{align}
where $K^{(G')}(x,y)$ is the kernel of the GUE with the boundary. 
However, as far as the zero- and one-instanton contribution is concerned, 
we can replace $K^{(G')}(x,y)$ with $K(x,y)$ in the standard GUE in \eqref{eq:K}. 
In fact, the kernel $K^{(G')}(x,y)$ in the GUE with the boundary is also given 
as in \eqref{eq:K}
\begin{align}
K^{(G')}(x,y)=&e^{-\frac{N}{4}\left(x^2+y^2\right)}\frac{1}{h_{N-1}^{(G')}}
\frac{p_N^{(G')}(x)p^{(G')}_{N-1}(y)-p^{(G')}_{N-1}(x)p_N^{(G')}(y)}{x-y},
\label{eq:G'K}
\end{align}
where $p_N^{(G')}$ is the orthogonal polynomial in the presence of the boundary 
\begin{align}
\int_{-\infty}^{\mu^2}dx\,e^{-\frac{N}{2}x^2}p_m^{(G')}(x)p_n^{(G')}(y)
=h_n^{(G')}\delta_{mn}. 
\end{align}
In \cite{Endres:2013sda} we explicitly demonstrate that differences are expanded as  
\begin{align}
&\tilde p_n(x)\equiv p_n^{(G')}(x)-p_n(x)=\tilde p_n^{(1)}(x)+\tilde p_n^{(2)}(x)+\cdots, \nn \\
&\tilde h_n\equiv h_n^{(G')}-h_n=\tilde h_n^{(1)}+\tilde h_n^{(2)}+\cdots
\end{align}
by taking account of the boundary effect iteratively and this expansion turns out to be 
in terms of the instanton number. Then difference of the kernel is written as 
\begin{align}
&\tilde K(x,y)\equiv K^{(G')}(x,y)-K(x,y)=\tilde K^{(1)}(x,y)+\tilde K^{(2)}(x,y)+\cdots, \nn \\
&\tilde K^{(1)}(x,y)\nn \\
&=e^{-\frac N4(x^2+y^2)}\frac{1}{x-y}\Biggl\{
\frac{1}{h_{N-1}}\left(\tilde L_N^{(1)}(x)+\tilde L_{N-1}^{(1)}(y)
-\frac{\tilde h_{N-1}^{(1)}}{h_{N-1}}\right)p_N(x)p_{N-1}(y)-(x\leftrightarrow y)\Biggr\},
\end{align}
where 
\begin{align}
\tilde L_n(x)\equiv\frac{p_n^{(G')}(x)}{p_n(x)}=\tilde L_n^{(1)}(x)+\tilde L_n^{(2)}(x)+\cdots.
\end{align}
Since later we will find that the second term in \eqref{eq:G'two-pt} 
is relevant for ambiguity in the one-instanton sector, we need to evaluate 
\begin{align}
\int_{-\infty}^{\mu^2} dxdy\,f(x)g(y)\tilde K^{(1)}(x,y)^2. 
\end{align}
By using the results in \cite{Endres:2013sda}, it is not difficult to see that the integrations 
on $x$ and $y$ are dominated by $x=y$ configuration. In fact, since it is shown in 
\cite{Endres:2013sda} that $\tilde L_N^{(1)}(x)=L(x,1)S_N^{(1)}$ 
where $L(x, 1)$ is a function of $\cO(N^0)$ and $S_N$ is an $N$-dependent constant,  
$x,y$-dependence of $K^{(1)}(x,y)$ is essentially the same. 
More precisely, $e^{-\frac{x^2}{4}}p_n(x)$ takes a form of 
$e^{-\frac{x^2}{4}}p_n(x)\propto\exp(-Nf_n(x))(1+\cO(1/N))$ with a function $f_n(x)$ 
of $\cO(N^0)$, and hence the saddle points of $x$ and $y$ integrations 
become the same in the large-$N$ limit, and as a consequence, 
we need to take $y\rightarrow x$ limit. 
Then the problem is reduced to the one-point function by \eqref{eq:RandK}.  
However, as mentioned in \cite{Kuroki:2019ets},  in the case of the one-point function, 
\begin{align}
\int_{-\infty}^{\mu^2}dx\,(\mu^2-x)^n\tilde\rho^{(1)}(x)
\end{align}
is shown to have $p\geq 2-$instanton weight. Thus in \eqref{eq:twp-ptNicolai} we restrict 
ourselves to up to the one-instanton sector, and by using \eqref{eq:G'two-pt} and replacing 
$K^{(G')}(x,y)$ with $K(x,y)$, we obtain  
\begin{align}
\left.\vev{\frac1N\tr\phi^p\frac1N\tr\phi^q}^{(1,0)}_c\right|_{\text{0-inst.}+\text{1-inst.}} 
=&
\left.\vev{\frac1N\tr(\mu^2-M)^{\frac{p+q}{2}}}^{(G')}\right|_{\text{0-inst.}+\text{1-inst.}}
 \nn \\
&-\frac{1}{N^2}\left.\int_{-\infty}^{\mu^2} dxdy\,
(\mu^2-x)^{\frac p2}(\mu^2-y)^{\frac q2}K(x,y)^2\right|_{\text{0-inst.}+\text{1-inst.}}.
\end{align}
When both $p$ and $q$ are odd, the first term on the right-hand side is the regular one-point  function 
and has no ambiguity. Hence we concentrate on the second term 
\begin{align}
G_{pq}\equiv\left.-\frac{1}{N^2}\int_{-\infty}^{\mu^2} dxdy\,
(\mu^2-x)^{\frac p2}(\mu^2-y)^{\frac q2}K(x,y)^2\right|_{\text{0-inst.}+\text{1-inst.}}
\label{eq:Gdef}
\end{align}
and examine whether it has ambiguity. 

Since we are interested in $G_{pq}$ in the double scaling limit, we set 
\begin{align}
\mu^2=2+N^{-\frac23}s, \quad x=2+N^{-\frac23}\xi, \quad y=2+N^{-\frac23}\eta. 
\label{eq:dslvariables}
\end{align}
This limit corresponds to the soft edge scaling limit in the random matrix theory, under which 
the kernel in the GUE becomes the Airy kernel: 
\begin{align}
&\lim_{N\rightarrow\infty}N^{-\frac23}K(2+N^{-\frac23}\xi,2+N^{-\frac23}\eta)
=K_{\Ai}(\xi,\eta), \nn \\
&K_{\Ai}(\xi,\eta)\equiv\frac{\Ai(\xi)\Ai'(\eta)-\Ai'(\xi)\Ai(\eta)}{\xi-\eta},
\label{eq:AiK}
\end{align}
and then $G_{pq}$ is given by  
\begin{align}
G_{pq}=-N^{-\frac{p+q}{3}-2}\int_{-\infty}^sd\xi d\eta\,(s-\xi)^{\frac p2}(s-\eta)^{\frac q2}
K_{\Ai}(\xi,\eta)^2.
\label{eq:Gafterdsl}
\end{align}
Let us consider $G_{pq}$ in the one-instanton sector. 
According to \eqref{eq:instantonsectors}, 
the integrations on $\lambda_i$ and $\lambda_j$ on the right-hand side in the first line 
in \eqref{eq:twp-ptNicolai} are now restricted as\footnote
{The integral from $b$ to $\infty$ is negligible in the double scaling limit.} 
\begin{align}
\int_0^a2\lambda_id\lambda_i\int_a^b2\lambda_jd\lambda_j\cdot
+\int_a^b2\lambda_id\lambda_i\int_0^a2\lambda_jd\lambda_j\cdot,
\end{align}
which becomes via the Nicolai mapping $x=\mu^2-\lambda_i^2$, $y=\mu^2-\lambda_j^2$, 
and \eqref{eq:dslvariables}, 
\begin{align}
N^{-\frac43}\int_0^sd\xi\int_{-\infty}^0d\eta\,\cdot
+N^{-\frac43}\int_{-\infty}^0d\xi\int_0^sd\eta\,\cdot.
\end{align}
Namely one of the integrations are in the perturbative region and the other in the nonperturbative region. 
Therefore,
\begin{align}
\left.N^{\frac{p+q}{3}+2}G_{pq}\right|_{\text{1-inst.}}=&
-\left(\int_0^sd\xi\int_{-\infty}^0d\eta
+\int_{-\infty}^0d\xi\int_0^sd\eta\right)(s-\xi)^{\frac p2}(s-\eta)^{\frac q2}
K_{\Ai}(\xi,\eta)^2 \nn \\
=&-\int_0^sd\xi\int_{-\infty}^0d\eta\,(s-\xi)^{\frac p2}(s-\eta)^{\frac q2}K_{\Ai}(\xi,\eta)^2
+(p\leftrightarrow q).
\label{eq:GpqbyKAi}
\end{align}
Let us consider the first term where $\xi\in[0,s]$ and $\eta\in(-\infty,0]$. 
Since later it turns out that relevant contribution comes from $\xi\sim s\gg 1$ 
and $|\eta|\gg 1$, we use the asymptotic form of the Airy function in the Airy kernel
\begin{align}
&\Ai(\xi)\sim\frac{e^{-z}}{2\sqrt{\pi}\xi^{\frac14}}\left(u_e(z)+u_o(z)\right) 
\qquad (\xi\gg 1), 
\label{eq:Aiiinf} \\
&\Ai(\eta)\sim\frac{1}{\sqrt{\pi}|\eta|^{\frac14}}
\left(\cos\left(w-\frac{\pi}{4}\right)u_e(w)+\sin\left(w-\frac{\pi}{4}\right)u_o(w)\right)
\qquad (-\eta\gg 1),
\label{eq:Ai-inf}
\end{align}
where 
\begin{align}
z\equiv\frac23\xi^{\frac32}, \qquad w\equiv\frac23|\eta|^{\frac32},
\end{align}
and
\begin{align}
&u_e(z)=\sum_{k=0}^{\infty}(-1)^k\frac{u_{2k}}{z^{2k}}, \qquad 
u_o(z)=\sum_{k=0}^{\infty}(-1)^k\frac{u_{2k+1}}{z^{2k+1}}, \nn \\
&u_k=\frac{(2k+1)(2k+3)\cdots(6k-1)}{216^k\,k!}.
\label{eq:udef}
\end{align} 
Plugging these into $K_{\Ai}(\xi,\eta)$ in \eqref{eq:AiK}, it is rewritten 
for $\xi\gg1$, $-\eta\gg1$ as 
\begin{align}
(\xi-\eta)K_{\Ai}(\xi,\eta) \nn \\
=\frac{e^{-z}}{2\pi\xi^{\frac14}|\eta|^{\frac14}}
\Biggl(
&\sin\left(w-\frac{\pi}{4}\right)
\left(|\eta|^{\frac12}(u_e(z)+u_o(z))v_e(w)+\xi^{\frac12}(v_e(z)+v_o(z))u_o(w)\right) \nn \\
+&\cos\left(w-\frac{\pi}{4}\right)
\left(-|\eta|^{\frac12}(u_e(z)+u_o(z))v_o(w)+\xi^{\frac12}(v_e(z)+v_o(z))u_e(w)\right)
\Biggr),
\label{eq:KAibyu}
\end{align}
where 
\begin{align}
&v_e(z)=\sum_{k=0}^{\infty}(-1)^k\frac{v_{2k}}{z^{2k}}, \qquad 
v_o(z)=\sum_{k=0}^{\infty}(-1)^k\frac{v_{2k+1}}{z^{2k+1}}, \nn \\
&v_k=\frac{6k+1}{1-6k}u_k.
\label{eq:vdef}
\end{align}

%%%%%%%%%%%%%%%%%%%%%%%%%%%%%%%%%%%%%%%%
\subsection{Saddle point method}
In the presence of $e^{-z}$ in \eqref{eq:KAibyu}, we first apply the saddle point method 
to the integration on $\xi$ in the first term in \eqref{eq:GpqbyKAi}. 
We define $f(\xi,\eta)$ by 
\begin{align}
\left.N^{\frac{p+q}{3}+2}G_{pq}\right|_{\text{1-inst.}}
\equiv -\int_0^sd\xi\int_{-\infty}^0d\eta\,e^{-f(\xi,\eta)}+(p\leftrightarrow q),
\label{eq:fdef}
\end{align}
then
\begin{align}
f(\xi,\eta)=&2z-\frac{p}{2}\ln(s-\xi)-\frac{q}{2}\ln(s-\eta)+2\ln(\xi-\eta)
-\frac12\ln\xi+\frac12\ln|\eta|+2\ln(2\pi) \nn \\
&-2\ln\Biggl\{
\left(\frac{|\eta|}{\xi}\right)^{\frac12}
\left(\sin\left(w-\frac{\pi}{4}\right)
-\frac32\cos\left(w-\frac{\pi}{4}\right)v_1|\eta|^{-\frac32}\right) \nn \\
&\phantom{-2\ln\Biggl\{\left(\frac{|\eta|}{\xi}\right)^{\frac12}}
+\cos\left(w-\frac{\pi}{4}\right)
+\frac32\sin\left(w-\frac{\pi}{4}\right)u_1|\eta|^{-\frac32}\Biggr\} \nn \\
&\times\left(1+\cO(\xi^{-\frac32})\right)\left(1+\cO(\eta^{-3})\right). 
\label{eq:fform}
\end{align}
We will find later that $|\eta|$ becomes $\cO(s)$ and hence we have to retain 
$\left(\frac{|\eta|}{\xi}\right)^{\frac12}$ term. 
{}From this definition,  a saddle point with respect to $\xi$: $\partial_{\xi}f(\xi_*,\eta)=0$ 
is near $\xi=s$ as 
\begin{align}
\xi_*=s+\frac{p}{4s^{\frac12}}+\cO(s^{-2}).
\label{eq:xi*}
\end{align}
This justifies the use of the asymptotic formula of the Airy function for $\xi\gg1$ 
in \eqref{eq:Aiiinf}. Here we note that since $\eta\in(-\infty,0]$, 
$\frac{1}{\xi_*-\eta}\leq\frac{1}{\xi_*}$ and it is at most of $\cO(s^{-1})$. 
We also have 
\begin{align}
&\partial_{\xi}^2f(\xi_*,\eta)=\frac{8s}{p}+\frac{3(p+4)}{p}s^{-\frac12}+\cO(s^{-2}), \nn \\
&\partial_{\xi}^nf(\xi_*,\eta)=\frac{p}{2}\Gamma(n)\left(-\frac{4s^{\frac12}}{p}\right)^n
\left(1+\frac{n(p+6)}{8}s^{-\frac32}+\cO(s^{-3})\right) \qquad (n\geq 3).
\label{eq:derivatives}
\end{align}
We recognize here that even if $\partial_{\xi}^2f(\xi_*,\eta)$ is of $\cO(s)$, 
we have to take account of all order in the Gaussian approximation 
because $\partial_{\xi}^nf(\xi_*,\eta)$ is of $\cO(s^{\frac{n}{2}})$. 
Using these equations, the Taylor expansion of $f(\xi,\eta)$ around $\xi=\xi_*$ reads 
\begin{align}
&f(\xi,\eta)=f(\xi_*,\eta)+\frac12\partial_{\xi}^2f(\xi_*,\eta)(\xi-\xi_*)^2
+\sum_{n=3}^{\infty}\frac{1}{n!}\partial_{\xi}^nf(\xi_*,\eta)(\xi-\xi_*)^n \nn \\
&=f(\xi_*,\eta)+2s^{\frac12}(\xi-\xi_*)+\frac{3(p+4)}{2p}s^{-\frac12}(\xi-\xi_*)^2
-\frac{p}{2}\ln\left(1+\frac{4s^{\frac12}}{p}(\xi-\xi_*)\right)+\cO(s^{-2}).
\end{align}
By setting 
\begin{align}
t=2s^{\frac12}(\xi-\xi_*),
\label{eq:tdef}
\end{align}
the integration on $\xi$ becomes 
\begin{align}
\int d\xi\,e^{-f(\xi,\eta)}
=e^{-f(\xi_*,\eta)}\frac{1}{2s^{\frac12}}\int dt\,
e^{-t}\left(1+\frac{2t}{p}\right)^{\frac p2}(1+\cO(s^{-\frac32})). 
\end{align}
Now let us consider the integration contour. In performing the integration on $\xi$ in \eqref{eq:fdef}, 
we find that the saddle point in \eqref{eq:xi*} is not in the integration region. 
Here we follow the prescription proposed in \cite{Kuroki:2019ets}. 
Namely, we rotate the integration contour $[0,s] $ by $\pm\pi$ 
around the branch point $\xi=s$ so that it will pass through the saddle point $\xi=\xi_*$ 
without going through any singularity. 
More precisely, in order to avoid the cut $[s,\infty)$ of $\ln(s-\xi)$ in \eqref{eq:fform}, 
we have to make the rotation by $\pm\pi$ for $s\rightarrow s\pm i\epsilon$ with $\epsilon>0$. 
Thus the contour becomes a one on the positive real axis 
in the opposite direction decreasing $\xi$, 
and terminating at $\xi=s$. From the definition of $t$ in \eqref{eq:tdef}, 
we zoom in the vicinity of the saddle point in the large-$s$ regime 
and hence the lower limit of the integration (after $\pm\pi$ rotation) is $+\infty$ 
as usual in the standard saddle point method. On the other hand, even for the variable $t$, 
the upper edge $\xi=s$ remains finite $t=-\frac p2+\cO(s^{-\frac32})$ 
due to $\eqref{eq:xi*}$ and \eqref{eq:tdef}. 
Thus in this prescription the $\xi$ integration can be performed as 
\begin{align}
\int d\xi\,e^{-f(\xi,\eta)}
=&e^{-f(\xi_*,\eta)}\frac{1}{2s^{\frac12}}\int_{\infty}^{-\frac p2} dt\,
e^{-t}\left(1+\frac{2t}{p}\right)^{\frac p2}(1+\cO(s^{-\frac32})) \nn \\
=&-e^{-f(\xi_*,\eta)}\left(\frac2p\right)^{\frac p2}e^{\frac p2}
\frac{1}{2s^{\frac12}}\Gamma\left(\frac p2+1\right)(1+\cO(s^{-\frac32})).
\label{eq:Gaussint}
\end{align}
The steepest descent method by choosing the contour passing through 
the saddle point in this way should provide the trans-series in the one-instanton sector. 
In fact, in \cite{Kuroki:2019ets} it is shown that in the case of the one-point function, 
this prescription works and we can check explicitly the resurgence under it. 
It should be noticed that as mentioned in \cite{Kuroki:2019ets}, the rotations of the contour 
by $\pm\pi$ according to $s\rightarrow s\pm i\epsilon$ give the same integrand and 
do not cause any difference. 
Thus so far there is no ambiguity between $s\rightarrow s\pm i\epsilon$. 
It is in fact the saddle point value $f(\xi_*,\eta)$ that makes difference. 
This situation is also the same as in the case of the one-point function \cite{Kuroki:2019ets}. 
The origin of ambiguity is, under $s\to s\pm i\epsilon$, 
\begin{align}
\ln(s-\xi_*) \rightarrow \ln(s\pm i\epsilon-\xi_*)=
\ln(\xi_*-s)\pm i\pi. 
\label{eq:origin}
\end{align}
Plugging the saddle point $\eqref{eq:xi*}$ into \eqref{eq:fform} 
and using \eqref{eq:Gaussint} and \eqref{eq:origin}, we get
\begin{align}
&\left.N^{\frac{p+q}{3}+2}G_{pq}\right|_{\text{1-inst.}} \nn \\
&=e^{\pm\frac p2\pi i}
\frac{1}{2^{\frac p2+3}\pi^2}\Gamma\left(\frac p2+1\right)
s^{-\frac p4}e^{-\frac43s^{\frac32}}(1+\cO(s^{-\frac32})) \nn \\
&\times
\int_{-\infty}^0d\eta\,\frac{1}{|\eta|^{\frac12}}(s-\eta)^{\frac q2-2}
\Bigg\{
|\eta|^{\frac12}s^{-\frac12}\left(\sin\left(w-\frac{\pi}{4}\right)
-\frac32\cos\left(w-\frac{\pi}{4}\right)v_1|\eta|^{-\frac32}\right) \nn \\
&\phantom{\times\int_{-\infty}^0d\eta\,\frac{1}{|\eta|^{\frac12}}(s-\eta)^{\frac q2-2}
|\eta|^{\frac12}s^{-\frac12}}
+\left(\cos\left(w-\frac{\pi}{4}\right)
+\frac32\sin\left(w-\frac{\pi}{4}\right)u_1|\eta|^{-\frac32}\right)
\Biggr\}^2(1+\cO(\eta^{-3})) \nn \\
&+(p\leftrightarrow q).
\label{eq:afterxiint}
\end{align}

%%%%%%%%%%%%%%%%%%%%%%%%%%%%%%%%%%%%%%%%
\subsection{Contribution from perturbative region}
Finally let us consider the $\eta$-integration. From \eqref{eq:afterxiint}, it reads 
\begin{align}
I_{\eta}&\equiv\int_0^{\infty}d\eta\,\frac{1}{\eta^{\frac12}}(s+\eta)^{\frac q2-2} \nn \\
&\times\biggl\{\sin\left(w-\frac{\pi}{4}\right)
\left(\eta^{\frac12}s^{-\frac12}+\frac32u_1\eta^{-\frac32}\right)
+\cos\left(w-\frac{\pi}{4}\right)\left(1-\frac32v_1s^{-\frac12}\eta^{-1}\right)\biggr\}^2
(1+\cO(\eta^{-3})). 
\end{align}
Expansion of the curly braces yields both oscillating terms with $\sin(2w)$, $\cos(2w)$, 
and non-oscillating one. Since $\eta\gg1$ contribution is relevant for the integration, 
we anticipate that the former ones oscillate quite rapidly and their integrals vanish. 
{}For this reason we assume that they do not contribute and simply drop them. 
In fact, this prescription enables us to compute the one-point function 
at higher genus via the Airy kernel $K_{\Ai}(\xi,\xi)$, which exactly 
reproduces the result derived by another method in \cite{Kuroki:2016ucm}.\footnote
{We thank F. Sugino for pointing out this fact.}
The prescription becomes necessary because we take the double scaling limit 
at the level of the integrand from \eqref{eq:Gdef} to \eqref{eq:Gafterdsl}. 
Originally the kernel consists of the orthogonal polynomials as in \eqref{eq:G'K} 
and in the double scaling limit they have the oscillating behavior. 
If we were able to take the double scaling limit after the integration on $x$ and $y$ 
in \eqref{eq:Gdef}, we would not have to make such an assumption.\footnote
{This is the reason why we do not compute the two-point function 
in the zero-instanton sector via the kernel.} 
Then the integration on $\eta$ are simplified as 
\begin{align}
I_{\eta}=\frac12\int_0^{\infty}d\eta\,\left(
\frac 1s\eta^{\frac12}(s+\eta)^{\frac q2-2}+\eta^{-\frac12}(s+\eta)^{\frac q2-2}
\right)\left(1+\cO(s^{-\frac12}\eta^{-1},\eta^{-\frac32})\right).
\end{align}
Since we are now computing the integration with respect to the one variable 
in the perturbative region, it is natural to expect that it is related to a quantity  
in the one-point function. In fact, from the definition \eqref{eq:jmn}, 
it is easy to see that the integrals above can be written in terms of $I_{m,n}$ 
as, for odd $m$, 
\begin{align}
\int_0^{\infty}d\xi\,\xi^{\frac m2}(s+\xi)^{\frac n2}
=\left(N^{-\frac23}\right)^{-\frac{m+n}{3}-1}\pi i\left(\frac i2\right)^mI_{m,n}.
\end{align} 
Hence from \eqref{eq:Iresult}, 
\begin{align}
\left.I_{\eta}\right|_{\text{univ.}}=\frac12\left(s^{-1}I_{1,q-4}+I_{-1,q-4}\right)
=-\frac{1}{2\sqrt{\pi}}
\frac{\Gamma\left(\frac q2\right)}{\Gamma\left(\frac{q+1}{2}\right)}
s^{\frac{q-3}{2}}\ln s\,(1+\cO(s^{-\frac32})).
\end{align}
Now it becomes clear why we keep $\left(\frac{|\eta|}{\xi}\right)^{\frac12}$ term 
in \eqref{eq:fform}. In fact, the saddle point of $\xi$ is of $\cO(s)$ as in \eqref{eq:xi*}, 
while the power of $\eta$ increases the power of $s$ according to \eqref{eq:Iresult}. 
Substituting this for \eqref{eq:afterxiint}, we find 
\begin{align}
\left.N^{\frac{p+q}{3}+2}G_{pq}\right|_{\text{1-inst., univ.}}
=&-e^{\pm\frac p2\pi i}
\frac{1}{2^{\frac p2+4}\pi^{\frac52}}\Gamma\left(\frac p2+1\right)
\frac{\Gamma\left(\frac q2\right)}{\Gamma\left(\frac {q+1}{2}\right)}
s^{-\frac p4+\frac q2-\frac32}e^{-\frac43s^{\frac32}}\ln s\,(1+\cO(s^{-\frac32})) \nn \\
&+(p\leftrightarrow q).
\end{align}
Thus if we assume $p\leq q$ as in \eqref{eq:0-instamb}, 
the first term is more leading on the right-hand side in this equation. 
Therefore, the ambiguity in the two-point function in the one-instanton sector is 
\begin{align}
&\text{Amb.}\,N^{\frac{p+q}{3}+2}
\left.\vev{\frac1N\tr\phi^p\frac1N\tr\phi^q}_c^{(1,0)}
\right|_{\text{1-inst., univ.}} \nn \\
&=-i(1+\delta_{pq})\sin\left(\frac{p}{2}\pi\right)\frac{1}{2^{\frac p2+3}\pi^{\frac52}}
\Gamma\left(\frac p2+1\right)
\frac{\Gamma\left(\frac q2\right)}{\Gamma\left(\frac {q+1}{2}\right)}
s^{-\frac{p}{4}+\frac{q}{2}-\frac32}e^{-\frac43s^{\frac32}}\ln s
\left(1+\cO(s^{-\frac32})\right), 
\label{eq:1-instamb}
\end{align}
which exactly cancels that in the zero-instanton sector given in \eqref{eq:0-instamb}. 
Thus we have confirmed that resurgence works at the leading order 
in the large-$s$ regime in the double scaling limit. 
This cancellation strongly supports validity of our prescription in \eqref{eq:Gaussint} 
rotating the contour by $\pm\pi$. 
It is also desirable to give more justification of this prescription mathematically 
via resurgence theory applied to an interval, or Lefschetz thimbles.     

Finally, concerning our motivation mentioned in Introduction, we make a comment 
on a relation to physics, in particular spontaneous SUSY breaking. 
As shown in \cite{Endres:2013sda}, it is triggered by the instanton 
in the supersymmetric double-well matrix model. As mentioned in \eqref{eq:Z_pinst_dsl}, 
its weight is proportional to $e^{-\frac43s^{\frac32}}$. Ambiguities in the zero- 
and one-instanton sector given in \eqref{eq:0-instamb} and \eqref{eq:1-instamb} 
suggest that they originate from the instanton. Thus the results in this paper 
as well as the ambiguity in the one-point function found in \cite{Kuroki:2019ets} 
would reveal a counterpart of the instanton in the type IIA superstring theory. 
For example, the power of $s$ in front of the instanton weight would provide 
information on the number of the collective modes around it. We have derived 
the ambiguity not only in the one-point function in \cite{Kuroki:2019ets} 
but in the two-point function here and, therefore, it is expected that 
more detailed information on such a nonperturbative object causing SUSY breaking 
would be provided from our result.

%%%%%%%%%%%%%%%%%%%%%%%%%%%%%%%%%%%%%%%%%%%%%%%%%%%%%%%%%%%%%%%%%%%%%%%%%%%%%
\section{Conclusions and discussions}
\label{sec:discussion}
%%%%%%%%%%%%%%%%%%%%%%%%%%%%%%%%%%%%%%%%%%%%%%%%%%%%%%%%%%%%%%%%%%%%%%%%%%%%%
In this paper, we derived the two-point function of the odd operators 
in the zero- and one-instanton sector at the leading order in the large-$s$ expansion 
under the double scaling limit of the SUSY double-well matrix model. 
We found that the ambiguity arises from the Borel resummation 
in the zero-instanton sector, and from the saddle point value in the one-instanton sector. 
The form of the ambiguity is consistent with the weight of the instanton 
in the matrix model. We explicitly confirmed that 
the two ambiguities cancel each other and thus clarified resurgence structure. 
Together with the check of resurgence for the one-point function done 
in \cite{Kuroki:2019ets}, we have clarified resurgence structure 
of different quantities within the same model. This kind of study would provide 
some insight and be instructive for development of resurgence theory itself. 
For example, in our case, in the zero-instanton sector 
the stringy behavior of  $(2h)!$ growth and  the Borel non-summability as its consequence
follow from the quantities in the Gaussian matrix model like $C_{h,3h-1}$ 
in \eqref{eq:highestcomp} for the one-point function, and $\gamma_{h,j}$ 
in \eqref{eq:two-ptbyI2} for the two-point function. They are in common with 
correlation functions of the odd (non-SUSY) and even (SUSY) operators. 
Because the latter should be Borel summable, we deduce that their perturbative expansion 
should terminate at finite order or be alternating. In this way we can identify 
the origin of factorial growth and operator dependence by comparing resurgence structure 
of several quantities. In the one-instanton sector, since we have two integration variables, 
we find the hybrid of perturbative and nonperturbative saddles. This observation 
would be useful for future study of resurgence structure. 

In order to make direct connection between the SUSY breaking 
and resurgence structure, the correlation functions of the $\phi^2$-resolvent 
\eqref{eq:R2} would play an important role. In fact, by multiplying functions 
and integrating it, it yields correlation functions of both SUSY (even) 
and non-SUSY (odd) operators as in \eqref{eq:byresolvent}. 
This implies that these two kinds of correlation functions can be related 
through  those of the $\phi^2$-resolvent. Since the correlation functions 
of the even operators can be used as order parameters of SUSY breaking, 
this fact will be useful to try to make connection between the SUSY breaking 
and resurgence structure, which is one of the main motivations of this work 
as we mentioned in Introduction. 

We can consider several applications of the results in this paper. 
As shown in \eqref{eq:byresolvent}, all the correlation functions of $\phi$ 
can be deduced from those of the $\phi^2$-resolvent, which are mapped 
to those of the resolvent in the Gaussian matrix model as in \eqref{eq:resolventbyNicolai}. 
However, the latter is known to be written by the kernel as 
\begin{align}
R_k(x_1,\cdots,x_k)=\det_{i,j=1,\cdots,k}K(x_i,x_j), 
\end{align}
and in this paper we concretely present how to evaluate the kernel 
in the double scaling limit when $x_i$ and $x_j$ are in the perturbative region 
or in the nonperturbative region. Thus it is expected that we can compute 
multi-point functions at arbitrary genus by using the results in this paper 
as building blocks. 
Finally, in the context of the GUE, we explicitly divide the two-point function 
$\Gamma_h(x,y)$ in \eqref{eq:two-ptandGamma} by $(x-y)^2$ 
as in \eqref{eq:smallgamma}, by which the two integrations in the two-point function 
can be separated and it can be rewritten as the sum of the products 
of the one-point functions.  
Although our result of the quotient in Appendix \ref{app:gamma} is restricted 
to the leading order in the soft edge scaling limit, it would be quite useful 
for computation of multi-point functions in several models 
in which the Nicolai mapping is available.

%%%%%%%%%%%%%%%%%%%%%%%%%%%%%%%%%%%%%%%%%%%%%%%%%%%%%%%%%%%%%%%%%%
\section*{Acknowledgements}
%%%%%%%%%%%%%%%%%%%%%%%%%%%%%%%%%%%%%%%%%%%%%%%%%%%%%%%%%%%%%%%%%%
We are grateful to Fumihiko Sugino for collaboration at an early stage of this work. 
We would like to thank Tatsuhiro Misumi and  Shinsuke Nishigaki for useful discussions 
and comments. 
The work of T.~K. is supported in part by a Grant-in-Aid 
for Scientific Research (C), 16K05335, 19K03834.

%%%%%%%%%%%%%%%%%%%%%%%%%%%%%%%%%%%%%%%%%%%%%%%%%%%%%%%%%%%%%%%%%%%%%%%%%%
%%%%%%%%%%%%%%%%%%%%% Appendix %%%%%%%%%%%%%%%%%%%%%%%%%%%%%%%%%%%%%%%%%%%%
%%%%%%%%%%%%%%%%%%%%%%%%%%%%%%%%%%%%%%%%%%%%%%%%%%%%%%%%%%%%%%%%%%%%%%%%%%
\appendix
%%%%%%%%%%%%%%%%%%%%%%%%%%%%%%%%%%%%%%%%%%%%%%%%%%%%%%%%%%%%%%%%%%%%%%%%%%%
\section{Derivation of $\gamma_{h,j}$}
\label{app:gamma}
\setcounter{equation}{0}
%%%%%%%%%%%%%%%%%%%%%%%%%%%%%%%%%%%%%%%%%%%%%%%%%%%%
In this appendix, we give an explicit form 
$\gamma_{h,j}\equiv\gamma_{h,j \,3h-1-j}$ defined in \eqref{eq:smallgamma}. 
Throughout this appendix, we fix the genus $h$ 
and abbreviate $\gamma_{h,j}$ to $\gamma_j$. Hence $\gamma_{j}=\gamma_{3h-1-j}$. 

{}Eq. \eqref{eq:smallgamma} reads for $h\in\bm N$
\begin{align}
&2\sum_{j=0}^hC_{j,3j-1}C_{h-j,3(h-j)-1}\xi^{3h-3j}\zeta^{3j} 
\Bigl((6j+1)(6(h-j)-1)\xi+(6j-1)(6(h-j)+1)\zeta\Bigr) \nn \\
&-16\sum_{j=0}^{h-1}C_{j,3j-1}C_{h-1-j,3(h-1-j)-1}\xi^{3(h-1-j)+2}\zeta^{3j+2} 
(36j^2-1)(36(h-1-j)^2-1) \nn \\
&=(\xi-\zeta)^2\sum_{j=0}^{3h-1}\gamma_{j}\xi^{3h-1-j}\zeta^{j}
=\sum_{j=0}^{3h-1}\gamma_j
\left(\xi^{3h+1-j}\zeta^j-2\xi^{3h-j}\zeta^{j+1}+\xi^{3h-1-j}\zeta^{j+2}\right). 
\label{eq:smallgammaexp}
\end{align}
Comparing each order in both sides, we have
\begin{align}
&{\cO}\left(\xi^{3(h-j)+1}\zeta^{3j}\right): \nn \\
&2(6j+1)(6(h-j)-1)C_{j,3j-1}C_{h-j,3(h-j)-1}=\gamma_{3j}-2\gamma_{3j-1}+\gamma_{3j-2}
\qquad (1\leq j\leq h-1), \label{gamma3jrecursion} \\
&{\cO}\left(\xi^{3(h-j)}\zeta^{3j+1}\right): \nn \\
&2(6j-1)(6(h-j)+1)C_{j,3j-1}C_{h-j,3(h-j)-1}=\gamma_{3j+1}-2\gamma_{3j}+\gamma_{3j-1}
\qquad (1\leq j\leq h-1), \label{gamma3j+1recursion} \\
&{\cO}\left(\xi^{3(h-j)-1}\zeta^{3j+2}\right): \nn \\
&-16(36j^2-1)(36(h-1-j)^2-1)C_{j,3j-1}C_{h-1-j,3(h-1-j)-1}
=\gamma_{3j+2}-2\gamma_{3j+1}+\gamma_{3j} \nn \\
&\phantom{2(6j+1)(6(h-j)-1)C_{j,3j-1}C_{h-j,3(h-j)-1}
=\gamma_{3j}-2\gamma_{3j-1}+\gamma_{3j-2}\qquad} (0\leq j\leq h-1). 
\label{gamma3j+2recursion}
\end{align}
We also find from ${\cO}\left(\xi^{3h+1}\right)$, ${\cO}\left(\xi^{3h}\zeta\right)$, 
${\cO}\left(\xi\zeta^{3h}\right)$, and ${\cO}\left(\zeta^{3h+1}\right)$ that 
\begin{align}
&\gamma_0=\gamma_{3h-1}=(1-6h)C_{h,3h-1}
=-\frac{1}{2\sqrt{\pi}}\left(\frac{16}{3}\right)^h
\frac{\Gamma\left(3h+\frac12\right)}{h!}, 
\label{eq:gamma0}\\
&\gamma_1=\gamma_{3h-2}=(3-6h)C_{h,3h-1}
=-\frac{3}{4\sqrt{\pi}}\left(\frac{16}{3}\right)^h
\frac{\Gamma\left(3h-\frac12\right)}{h!}(2h-1), 
\label{eq:gamma1}
\end{align}
where we have used \eqref{eq:highestcomp}. 
{}From these initial values, we can determine $\gamma_j$ iteratively. 
For example, using \eqref{gamma3j+2recursion}, 
\begin{align}
\gamma_2=&-\gamma_0+2\gamma_1-8\left(36(h-1)^2-1\right)C_{h-1,3(h-1)-1} \nn \\
=&-\frac{3}{8\sqrt{\pi}}\left(\frac{16}{3}\right)^h
\frac{\Gamma\left(3h-\frac32\right)}{h!}(12h^2-12h+5),
\label{eq:gamma2}
\end{align}
and by \eqref{gamma3jrecursion}, 
\begin{align}
\gamma_3=&-\gamma_1+2\gamma_2+14C_{h-1,3(h-1)-1}(6(h-1)-1) \nn \\
=&-\frac{1}{\sqrt{\pi}}\left(\frac{16}{3}\right)^{h-1}
\frac{\Gamma\left(3h-\frac52\right)}{h!}(h-1)(72h^2-60h+35).
\label{eq:gamma3}
\end{align}
Setting $\delta_j=\gamma_j-\gamma_{j-1}$ ($j\in\bm N$), 
\eqref{eq:highestcomp} and \eqref{gamma3jrecursion} leads to 
\begin{align}
\delta_{3j}-\delta_{3j-1}
&=\left(\gamma_{3j}-\gamma_{3j-1}\right)-\left(\gamma_{3j-1}-\gamma_{3j-2}\right) 
 \nn \\
&=\frac{1}{2\pi}\left(\frac{16}{3}\right)^h
\frac{\Gamma\left(3j-\frac12\right)}{j!}\frac{\Gamma\left(3(h-j)+\frac12\right)}{(h-j)!}
\left(3j+\frac12\right) \qquad  (1\leq j\leq h-1). 
\label{eq:delta3jdef}
\end{align}
Similarly, from \eqref{gamma3j+1recursion} and \eqref{gamma3j+2recursion},  
\begin{align}
&\delta_{3j+1}-\delta_{3j}
=\frac{1}{2\pi}\left(\frac{16}{3}\right)^h
\frac{\Gamma\left(3j+\frac12\right)}{j!}\frac{\Gamma\left(3(h-j)-\frac12\right)}{(h-j)!}
\left(3(h-j)+\frac12\right) \qquad  (1\leq j\leq h-1), 
\label{eq:delta3j+1def}\\
&\delta_{3j+2}-\delta_{3j+1}
=-\frac{3}{\pi}\left(\frac{16}{3}\right)^h
\frac{\Gamma\left(3j+\frac32\right)}{j!}\frac{\Gamma\left(3(h-j)-\frac32\right)}{(h-1-j)!}
\qquad  (0\leq j\leq h-1).
\label{eq:delta3j+2def}
\end{align}
Therefore, for $2\leq j\leq h-1$,
\begin{align}
\delta_{3j}=&\sum_{k=2}^j
\left(\delta_{3k}-\delta_{3k-1}+\delta_{3k-1}-\delta_{3k-2}
+\delta_{3k-2}-\delta_{3k-3}\right)+\delta_3 \nn \\
=&-\frac{1}{2\pi}\left(\frac{16}{3}\right)^h\frac{1}{h}\frac{h-2j}{j!(h-j)!}
\Gamma\left(3j+\frac12\right)\Gamma\left(3(h-j)+\frac12\right),
\label{eq:delta3j}
\end{align}
where we have utilized \eqref{eq:gamma2}, \eqref{eq:gamma3}, 
\eqref{eq:delta3jdef}, \eqref{eq:delta3j+1def} and \eqref{eq:delta3j+2def}, 
and taken the sum on $k$. It is easy to check that this equation holds 
for $1\leq j\leq h-1$. Likewise, we obtain 
\begin{align}
\delta_{3j+1}=&\sum_{k=1}^j
\left(\delta_{3k+1}-\delta_{3k}+\delta_{3k}-\delta_{3k-1}
+\delta_{3k-1}-\delta_{3k-2}\right)+\delta_1 \nn \\
=&\frac{1}{\pi}\left(\frac{16}{3}\right)^h
\frac{\Gamma\left(3j+\frac32\right)\Gamma\left(3(h-j)-\frac12\right)}{hj!(h-1-j)!}, 
\label{eq:delta3j+1} \\
\delta_{3j+2}=&\sum_{k=1}^j
\left(\delta_{3k+2}-\delta_{3k+1}+\delta_{3k+1}-\delta_{3k}
+\delta_{3k}-\delta_{3k-1}\right)+\delta_2 \nn \\
=&-\frac{1}{\pi}\left(\frac{16}{3}\right)^h
\frac{\Gamma\left(3j+\frac52\right)\Gamma\left(3(h-j)-\frac32\right)}{hj!(h-1-j)!} 
\label{eq:delta3j+2} 
\end{align} 
for $0\leq j\leq h-1$. Hence 
\begin{align}
\gamma_{3j}=\sum_{k=1}^j\left(\delta_{3k}+\delta_{3k-1}+\delta_{3k-2}\right)+\gamma_0
\end{align}
for $1\leq j\leq h-1$ and each sum yields 
\begin{align}
\gamma_{3j}^{(1)}\equiv&
\sum_{k=1}^j\delta_{3k}+\gamma_0 \nn \\
=&\frac{1}{\pi}\left(\frac{16}{3}\right)^h\Biggl[
\frac{1}{2h}\frac{\Gamma\left(3j+\frac72\right)}{(j+1)!}
\frac{\Gamma\left(3(h-j)-\frac52\right)}{(h-j-1)!} \nn \\
&\times\biggr\{
-(j+1){}_5F_4\biggl(1,j+\frac76,j+\frac32,j+\frac{11}{6},-h+j+1; \nn \\
&\phantom{\times\biggr\{-(j+1){}_5F_4\bigl(}
-h+j+\frac76,-h+j+\frac32,-h+j+\frac{11}{6},j+1; 1\biggr) \nn \\
&\phantom{\times\biggr\{}
+(h-j-1){}_5F_4\biggl(1,j+\frac76,j+\frac32,j+\frac{11}{6},-h+j+2; \nn \\
&\phantom{\times\biggr\{+(h-j-1){}_5F_4\bigl(}
-h+j+\frac76,-h+j+\frac32,-h+j+\frac{11}{6},j+2; 1\biggr)
\biggr\} \nn \\
&-\frac{\sqrt{\pi}}{2}\frac{\Gamma\left(3h+\frac12\right)}{h!}
{}_4F_3\left(\frac16,\frac12,\frac56,1-h;
\frac16-h,\frac12-h,\frac56-h; 1\right) \nn \\
&+\frac{15\sqrt{\pi}}{16}\frac{\Gamma\left(3h-\frac52\right)}{h!}
{}_4F_3\left(\frac76,\frac32,\frac{11}{6},1-h; \frac76-h,\frac32-h,\frac{11}{6}-h; 1\right)
\Biggr],
\label{eq:gamma3j1}
\end{align}
\begin{align}
\gamma_{3j}^{(2)}\equiv&
\sum_{k=1}^j\delta_{3k-1} \nn \\
=&\frac{1}{\pi}\left(\frac{16}{3}\right)^h\Biggl[
\frac{1}{h}\frac{\Gamma\left(3j+\frac52\right)}{j!}
\frac{\Gamma\left(3(h-j)-\frac32\right)}{(h-j-1)!} \nn \\
&\phantom{\frac{1}{\pi}\left(\frac{16}{3}\right)^h\Biggl[}
\times{}_5F_4\biggl(1,j+\frac56,j+\frac76,j+\frac32,-h+j+1; \nn \\
&\phantom{\frac{1}{\pi}\left(\frac{16}{3}\right)^h\Biggl[\times{}_5F_4\biggl(}
-h+j+\frac56,-h+j+\frac76,-h+j+\frac32,j+1; 1\biggr) \nn \\
&-\frac{3\sqrt{\pi}}{4}\frac{\Gamma\left(3h-\frac32\right)}{h!}
{}_4F_3\left(\frac56,\frac76,\frac32,1-h; \frac56-h,\frac76-h,\frac32-h; 1\right)
\Biggr],
\end{align}
\begin{align}
\gamma_{3j}^{(3)}\equiv&
\sum_{k=1}^j\delta_{3k-2} \nn \\
=&\frac{1}{\pi}\left(\frac{16}{3}\right)^h\Biggl[
-\frac{1}{h}\frac{\Gamma\left(3j+\frac32\right)}{j!}
\frac{\Gamma\left(3(h-j)-\frac12\right)}{(h-j-1)!} \nn \\
&\phantom{\frac{1}{\pi}\left(\frac{16}{3}\right)^h\Biggl[}
\times{}_5F_4\biggl(1,j+\frac12,j+\frac56,j+\frac76,-h+j+1; \nn \\
&\phantom{\frac{1}{\pi}\left(\frac{16}{3}\right)^h\Biggl[\times{}_5F_4\biggl(}
-h+j+\frac12,-h+j+\frac56,-h+j+\frac76,j+1; 1\biggr) \nn \\
&+\frac{\sqrt{\pi}}{2}\frac{\Gamma\left(3h-\frac12\right)}{h!}
{}_4F_3\left(\frac12,\frac56,\frac76,1-h; \frac12-h,\frac56-h,\frac76-h; 1\right)
\Biggr].
\end{align}
It turns out that \eqref{eq:gamma3j1} also holds for $j=0$. Using these results, 
we have obtained $\gamma_{j}$ ($0\leq j\leq 3h-1$) as 
\begin{align}
&\gamma_{3j}=\gamma_{3j}^{(1)}+\gamma_{3j}^{(2)}+\gamma_{3j}^{(3)}, \nn \\
&\gamma_{3j+1}=\gamma_{3j}+\delta_{3j+1}, \nn \\
&\gamma_{3j+2}=\gamma_{3j}+\delta_{3j+1}+\delta_{3j+2},
\end{align} 
where $\delta_{3j+1}$ and $\delta_{3j+2}$ are given in \eqref{eq:delta3j+1} and 
\eqref{eq:delta3j+2}, respectively. 
{}From the discussions in section \ref{sec:zero-inst},  we recognize that 
$\gamma_h(\xi,\eta)$ defined in \eqref{eq:smallgamma} by using $\gamma_j$'s are proportional 
to the leading term of the two-point function of the resolvent \eqref{eq:two-ptandGamma} 
divided by $(x-y)^2$ under the double scaling limit \eqref{eq:dsl}. 
This limit is the soft edge scaling limit of the random matrix theory 
and from $\Gamma_h(x,y)$ we derive any two-point function. 
Hence our result above would be quite useful in computation of the two-point functions 
in the random matrix theory because it makes the integrations over $x$ and $y$ decoupled 
and two independent ones.\footnote
{In \cite{Kuroki:2012nt},  we take another method to get rid 
of the denominator $(x-y)^2$, but it would be difficult to apply it 
to the higher genus case.}

%%%%%%%%%%%%%%%%%%%%%%%%%%%%%%%%%%%%%%%%%%%%%%%%%%%%%%%%%%%%%%%%%%%%%%%%%%%
\section{Properties of $\gamma_{h,j}$}
\label{app:gammaprop}
\setcounter{equation}{0}
%%%%%%%%%%%%%%%%%%%%%%%%%%%%%%%%%%%%%%%%%%%%%%%%%%%%
In this appendix, we prove some properties of $\gamma_{h,j}\equiv\gamma_{h,j \,3h-1-j}$ 
defined in \eqref{eq:smallgamma} which play important roles 
in derivation of ambiguity in the zero-instanton sector. 
In this appendix, we assume $j\in\bm Z_{\geq 0}$ unless otherwise specified. \\

\noindent
\underline{Proposition 1.} 

There exists a polynomial $f_j(h)$ of $h$ of degree $j$ satisfying 
\begin{align}
\gamma_{h,j}=\frac{\Gamma\left(3h+\frac12-j\right)}{h!}\left(\frac{16}{3}\right)^hf_j(h).
\end{align}

\noindent
\underline{Proof} 

We prove this by induction. Eqs.~\eqref{eq:gamma0}, \eqref{eq:gamma1}, 
and \eqref{eq:gamma2} imply that the statement holds for $j=0,1,2$. 
Suppose $f_j(h)$ exists for $j=3k, 3k+1, 3k+2$ ($k\in\bm Z_{\geq 0}$). 
Then by using \eqref{eq:delta3j}, we have
\begin{align}
\gamma_{h,3k+3}=&\gamma_{h,3k+2}+\delta_{3k+3} \nn \\
=&\frac{\Gamma\left(3h+\frac12-(3k+2)\right)}{h!}\left(\frac{16}{3}\right)^hf_{3k+2}(h) \nn \\
&-\frac{1}{2\pi}\left(\frac{16}{3}\right)^h\frac{1}{h}\frac{h-2(k+1)}{(k+1)!(h-(k+1))!}
\Gamma\left(3(k+1)+\frac12\right)\Gamma\left(3\left(h-(k+1)\right)+\frac12\right) \nn \\
=&\frac{\Gamma\left(3h-3k-\frac52\right)}{h!}\left(\frac{16}{3}\right)^h
\biggl\{\left(3h-3k-\frac52\right)f_{3k+2}(h) \nn \\
&-\frac{1}{2\pi}\left(h-2(k+1)\right)
\frac{(h-1)!}{(k+1)!(h-(k+1))!}\Gamma\left(3(k+1)+\frac12\right)
\biggr\}.
\end{align}
Here it is easy to see that the equation in the curly braces is a polynomial of $h$ of degree $3k+3$. 
Similarly, we can find that polynomials $f_{3k+4}(h)$ and $f_{3k+5}(h)$ exist. 
$\quad \blacksquare$ \\

\noindent
\underline{Proposition 2.} 
\begin{align}
f_j(h)=-\frac{3^j}{2\sqrt{\pi}}h^j+\frac{3^{j-1}}{4\sqrt{\pi}}(j^2+2)h^{j-1}+{\cO}(h^{j-2}). 
\end{align}

\noindent
\underline{Proof} 

In the proof of the previous proposition, we found that 
\begin{align}
\gamma_{h,j}=&\gamma_{h,j-1}+\delta_j, \label{eq:gammarecursion}\\
&\gamma_{h,j-1}=\left(\frac{16}{3}\right)^h
\frac{\Gamma\left(3h+\frac12-j\right)}{h!}\left(3h+\frac12-j\right)f_{j-1}(h), \nn \\
&\delta_j=\left(\frac{16}{3}\right)^h\frac{\Gamma\left(3h+\frac12-j\right)}{h!}
\left(\text{polynomial of degree}\left[\frac{j}{3}\right]\right). \nn
\end{align}
Thus when $j\geq 4$, in arguing $\cO(h^{j-1})$ and $\cO(h^{j-2})$ terms in $f_{j-1}(h)$, 
we can neglect the contribution from $\delta_j$. Setting 
\begin{align}
f_j(h)=c_j^{(j)}h^j+c_{j-1}^{(j)}h^{j-1}+\cO(h^{j-2}), 
\end{align}
and comparing the terms of $\cO(h^j)$ and $\cO(h^{j-1})$ 
in both sides in \eqref{eq:gammarecursion},  we get 
\begin{align}
&c_j^{(j)}=3c_{j-1}^{(j-1)}, \nn \\
&c_{j-1}^{(j)}=\left(\frac12-j\right)c_{j-1}^{(j-1)}+3c_{j-2}^{(j-1)} 
\label{eq:cjrecursion}
\end{align}
for $j\geq 4$. 
{}From eqs.~\eqref{eq:gamma0}, \eqref{eq:gamma1}, \eqref{eq:gamma2}, 
and \eqref{eq:gamma3}, we see that the first equation holds even for $j\in\bm N$ and that 
$c_0^{(0)}=-\frac{1}{2\sqrt{\pi}}$. Therefore, 
\begin{align}
c_j^{(j)}=-\frac{3^j}{2\sqrt{\pi}}. 
\end{align}
Substituting this for the second equation in \eqref{eq:cjrecursion} and solving it, we obtain
\begin{align}
c_{j-1}^{(j)}=\frac{3^{j-1}}{4\sqrt{\pi}}(j^2+2),
\end{align}
and it is easy to check that it is true for $j\in\bm N$. $\quad \blacksquare$ \\

\noindent
\underline{Proposition 3.} 
\begin{align}
f_j(h)\binomi{\alpha}{h}=\sum_{k=0}^jb_k^{(j)}(\alpha)\binom{\alpha-k}{h-k}, 
\end{align}
where $b_k^{(j)}(\alpha)$ ($0\leq k\leq j$) are independent of $h$, and 
\begin{align}
&b_j^{(j)}(\alpha)=c_j^{(j)}\frac{\Gamma(\alpha+1)}{\Gamma(\alpha+1-j)}
=-\frac{3^j}{2\sqrt{\pi}}\frac{\Gamma(\alpha+1)}{\Gamma(\alpha+1-j)}, \\
&b_{j-1}^{(j)}(\alpha)=-\frac{3^{j-1}}{4\sqrt{\pi}}(j-2)(2j+1)
\frac{\Gamma(\alpha+1)}{\Gamma(\alpha+2-j)} \qquad (j\in\bm N).
\end{align}

\noindent
\underline{Proof} 

We use the identity
\begin{align}
h(h-1)\cdots \left(h-(j-1)\right)\binom{\alpha}{h}
=\alpha(\alpha-1)\cdots\left(\alpha-(j-1)\right)\binom{\alpha-j}{h-j}.
\label{eq:binomialidentity}
\end{align}
Thus we rewrite $f_j(h)$ as 
\begin{align}
f_j(h)=&c_j^{(j)}h^j+c_{j-1}^{(j)}h^{j-1}+\cO(h^{j-2}) \nn \\
=&c_j^{(j)'}h(h-1)\cdots\left(h-(j-1)\right)
+c_{j-1}^{(j)'}h(h-1)\cdots\left(h-(j-2)\right)+\cO(h^{j-2}).
\end{align}
Then 
\begin{align}
c_j^{(j)}=c_j^{(j)'}, \qquad 
c_{j-1}^{(j)'}=c_{j-1}^{(j)}+\frac{j(j-1)}{2}c_j^{(j)'}
=-\frac{3^{j-1}}{4\sqrt{\pi}}(j-2)(2j+1),
\end{align}
and by \eqref{eq:binomialidentity}
\begin{align}
&f_j(h)\binom{\alpha}{h} \nn \\
&=c_j^{(j)'}h(h-1)\cdots\left(h-(j-1)\right)\binom{\alpha}{h}
+c_{j-1}^{(j)'}h(h-1)\cdots\left(h-(j-2)\right)\binom{\alpha}{h}
+\cO(h^{j-2})\binom{\alpha}{h} \nn \\
&=c_j^{(j)'}\frac{\Gamma(\alpha+1)}{\Gamma(\alpha+1-j)}\binom{\alpha-j}{h-j}
+c_{j-1}^{(j)'}\frac{\Gamma(\alpha+1)}{\Gamma(\alpha+2-j)}\binom{\alpha-j+1}{h-j+1}
+\sum_{k=0}^{j-2}b_k^{(j)}(\alpha)\binom{\alpha-k}{h-k},
\end{align}
where evidently $b_k^{(j)}(\alpha)$ ($0\leq k\leq j$) does not depend on $h$, and 
\begin{align}
&b_j^{(j)}(\alpha)=c_j^{(j)'}\frac{\Gamma(\alpha+1)}{\Gamma(\alpha+1-j)}
=-\frac{3^j}{2\sqrt{\pi}}\frac{\Gamma(\alpha+1)}{\Gamma(\alpha+1-j)},\nn \\
&b_{j-1}^{(j)}(\alpha)=c_{j-1}^{(j)'}\frac{\Gamma(\alpha+1)}{\Gamma(\alpha+2-j)}
=-\frac{3^{j-1}}{4\sqrt{\pi}}(j-2)(2j+1)
\frac{\Gamma(\alpha+1)}{\Gamma(\alpha+2-j)} \qquad (j\in\bm N).\quad \blacksquare
\end{align}

%%%%%%%%%%%%%%%%%%%%% References %%%%%%%%%%%%%%%%%%%%%%%%%%%%%%%%%%%%%%%

\end{document}